\documentclass[12pt]{article}
\usepackage{amssymb}
\usepackage{graphicx}
\usepackage[hang,small,bf]{caption}
\usepackage{hyperref}

\newcommand{\be}[3]{\begin{equation}  \label{#1#2#3}}
\newcommand{\ee}{\end{equation}}
\newcommand{\ba}{\begin{array}}
\newcommand{\ea}{\end{array}}
\newcommand{\bea}[3]{\begin{eqnarray}  \label{#1#2#3}}
\newcommand{\eea}{\end{eqnarray}}

\renewcommand{\arraystretch}{1.5}

\let\Large=\large
\let\large=\normalsize

\def\I{\mathbb I}

\def\1{\mathbb 1}

\def\N{$\cal N \, $}

\newcommand{\haken}{\mathbin{\hbox to 8pt{%
                 \vrule height0.4pt width7pt depth0pt
                 \kern-.4pt
                 \vrule height4pt width0.4pt depth0pt\hss}}}
\newcommand{\resetcounter}{\setcounter{equation}{0}}

\renewcommand{\theequation}{\thesection.\arabic{equation}}

\begin{document}

\thispagestyle{empty}


\DeclareGraphicsExtensions{.jpg,.pdf,.mps,.png}
\begin{flushright}
\baselineskip=18pt
UPR-1139-T, LMU-ASC 73/05, MPP-2005-161, hep-th/0512032 \\
\end{flushright}

\begin{center}
\vglue 1.5cm

{\Large\bf Classification of Supersymmetric Flux Vacua in M
Theory} \vglue 2.0cm {\Large Klaus Behrndt$^a$, Mirjam Cveti\v
c$^b$, and Tao Liu$^b$} \vglue 1cm {$^a$ Email:
behrndt@theorie.physik.uni-muenchen.de\\\vspace*{6pt}
Arnold-Sommerfeld-Center
for Theoretical Physics \\
 Department f\"ur Physik,
Ludwig-Maximilians-Universit\"at
M\"unchen, \\
Theresienstra\ss{}e 37, 80333 M\"unchen, Germany \\

\vspace{.2truecm}

and

\vspace{.2truecm}

Max-Planck-Institut f\"ur Physik,
F\"ohringer Ring 6, 80805 M\"unchen, Germany

\vspace*{18pt}

$^b$ Email: cvetic@cvetic.hep.upenn.edu\\
    liutao@physics.upenn.edu\\\vspace*{6pt}
Department of Physics and Astronomy,
University of Pennsylvania, \\Philadelphia, PA 19104-6396, USA \\
}
\end{center}



\begin{abstract}

\noindent We present a comprehensive classification of
supersymmetric vacua of M-theory compactification  on
seven-dimensional manifolds with general four-form fluxes. We
analyze the cases where the resulting four-dimensional vacua have
\N = 1,2,3,4 supersymmetry and the internal space allows for
$SU(2)$, $SU(3)$ or $G_2$ structures. In particular, we find for
\N = 2 supersymmetry, that the external space-time is Minkowski
and the base manifold of the internal space is conformally
K\"ahler for $SU(2)$ structures, while for $SU(3)$ structures the
internal space has to be Einstein-Sasaki  and  no internal fluxes
are allowed. Moreover, we provide a new vacuum with \N = 1
supersymmetry and $SU(3)$ structure, where all fluxes are non-zero
and the first order differential equations are solved.

\end{abstract}


\newpage

\tableofcontents

\newpage


\section{Introduction}
\resetcounter

One of the major problems confronting string compactification is
the emergence of a huge degeneracy of string vacua due to flat
directions in moduli space. Non-perturbative (D-brane) gauge
dynamics and background supergravity fluxes provide two promising,
dual approaches to lift such a vacuum degeneracy.  In this paper
we shall focus on the effects of background supergravity fluxes.
On one hand, the gravitational effects, induced by fluxes, can
expand or contract  cycles which are parallel or perpendicular to
fluxes and their competing effects may lead to stabilization of
moduli in the closed string sector. On the other hand, fluxes also
couple to the D-brane world-volume action, which in turn
introduces a non-trivial potential for the moduli in the open
string sector, thus providing stabilization of the open string
moduli. Note, both of these stabilization effects  are achieved at
the classical level.

Over the past years a significant progress has been made in our
understanding of vacua in the presence of background fluxes. In first
attempts Calabi-Yau compactifications in the presence of fluxes have
been considered \cite{9510227, 9610151,9911011,9912152,0001082,
0012213, 0202168}.  In supersymmetric vacua fluxes and geometry are
directly linked to each other and, in general, fluxes do not respect
the special holonomy of the internal space, because they generate a
back-reaction on the geometry. This deformation can be expressed by
non-zero torsion classes of the geometry (for a review see, e.g.,
\cite{0212008} and references therein). In the simplest case, the
back-reaction is given only by a non-trivial warp factor. Especially
on the type IIA string theory and M-theory side, fluxes generate a
severe back reaction on the internal geometry
\cite{0012152,0206213,0211102,0302047,0303127,0308045,0308202,
0411276,0411279,0412006} and only few examples are explicitly known
\cite{0403049,0407263,0412250}; for reviews we refer the reader to
\cite{DaAg,0503129,0509003} and references therein.

Since there are many supersymmetric vacua, it is important to
develop an explicit analysis that would provide a comprehensive
classification of such vacua. Each supersymmetric vacuum implies
the existence of a Killing spinor, which has to be a singlet under
the structure group of the underlying manifold. If there are no
fluxes, the Killing spinor is covariantly constant and hence the
holonomy has to be restricted and coincides with the structure
group. In this case, there is a one-to-one correspondence between
the holonomy and the amount of unbroken supersymmetry. But if the
fluxes do not vanish, the holonomy is generically not restricted
anymore and supersymmetric vacua are classified by the structure
group.  Also, there is not any longer a direct link between the
amount of unbroken supersymmetry and the structure group -- the
same group can give rise to vacua with different amount of
supersymmetry, as we will encounter below. Depending on the
geometry, certain fluxes can namely be added without breaking any
supersymmetry whereas others impose additional constraints. Since
the Killing spinor(s) have to be a singlet under the structure
group, the classification is equivalent to the number of
independent internal spinors, ie.\ the larger the structure group
the simpler the spinor Ansatz and the simpler the solution.

Having non-trivial Killing spinors, one can build differential forms
as fermionic bi-linears. These forms are singlets under the structure
group $G$ and satisfy algebraic constraints and first order
differential equations, which can be derived from the Killing spinor
equations and are known as $G$-structures \cite{0205050,m0202282}.
For specific cases, one can already infer constraints on the geometry
from the existence of these forms. It is e.g.\ well-known that a
6-dimensional spin manifold allows for regular vectors only if the
Euler number is zero. On the other hand, for a 7-dimensional spin
manifold one can always define three vectors, which implies that one
can, without making any constraint on the geometry, express the vacuum
in terms of SU(2) structures. In type II string theory, vacua with
$SU(2)$ structure have been discussed in
\cite{0308045,0403220,0506160}

Flux compactifications of M-theory with a vanishing cosmological
constant have been considered in \cite{9908088, 0010282, 0012194,
0303127} and compactifications to a 4-dimensional anti deSitter space
are discussed in
\cite{0007213,0302047,0308046,0311119,0403235,0411194,0502200}. The
amount of unbroken supersymmetry is related to the number of external
spinors which are either Weyl or Majorana. For an \N = 1 vacuum, the
spinor Ansatz has just one external spinor and this case has been
explored in most papers. Much less has been done for \N = 2 (see
however \cite{0506160}) or even \N = 4. As we will see below, these
cases are highly constraint.

The focus of this paper is on the study of four-dimensional
supersymmetric vacua of M-theory with a general 4-form flux, which has
components in the internal space as well as a Freund-Rubin
parameter. We do not require that the external space is flat, but
allow also for a non-vanishing (negative) cosmological constant. In
particular, we provide a systematic classification of four-dimensional
supersymmetric M-theory vacua by deriving and analyzing explicit
conditions that have at least $SU(2)$ structure (thus, also
encompassing $SU(3)$ and $G_2$ structures).  Hence, we consider the
most general case without putting any constraints on the geometry. The
core analysis is based on the constraints for fluxes and torsion
classes of the internal space that arise from the Killing spinor
equations. In addition, we also implement constrains that arise from
the Bianchi identities and the equations of motion for fluxes.

Strictly speaking, the Killing spinors have to be globally
well-defined, which would not be the case if in addition to the
background fluxes one takes brane configurations into account. On
the other hand, it may happen that the Killing spinor equations
have non-trivial solutions only if one introduces sources for the
fluxes (e.g.\ when expressed in terms of harmonic functions),
which are nothing but branes. Moreover, solving the Killing spinor
equations is a local analysis, but in order to have a consistent
vacuum one has to address also global issues. For example, the
volume should be finite and sufficiently small and there should be
no net-charge on the internal space. We will not address these
important issues here and restrict ourselves to a local analysis
of the background supergravity fluxes, only.

The plan of the paper is as follows. In the next Section we shall discuss in
detail the different structure groups and define the corresponding global
differential forms. In Section 3 we discuss the Ans\"atze for the bosonic
fields and the Killing spinors related to the different G-structures. We also
give the appropriate decomposition of fluxes.  With these spinors, we
investigate in Section 4, 5 and 6 the Killing spinor equations (as well as
Bianchi identities and equations of motions for fluxes) for  the cases where
the resulting four-dimensional vacua have \N = 4,3, \N=2 and \N=1
supersymmetry, respectively. In each case we derive the explicit constraints
for the fluxes and the torsion classes of the internal space. In Section 7 we
conclude with a summary of results for this comprehensive classification and
outline some directions for future investigations. In Appendix A provides
notational conventions for spinors and gamma matrices, and in Appendix B we
give additional explicit equations for \N = 2 $G$-structures.


\section{Group Structures}
\resetcounter

In this section we introduce group structures for different
groups. For our conventions and notation, we basically follow
\cite{0302047, 0311119, 0406138}.  We also refer the reader
to \cite{0212008} for  more detailed discussions.

\subsection{Defining $G$ - structures}


A convenient way to define $G$--structures is via $G$--invariant
spinors and tensors. By considering the set of orthonormal frames the
structure group of the frame bundle reduces to $G \subset O(d)$ or $G
\subset Spin(d)$ for spin manifold. Therefore, the existence of these
$G$--invariant spinors and tensors on a $d$--dimensional Riemannian
manifold implies a reduced structure group of the frame bundle.

Since they are the singlets of the reduced structure group, these
$G$--invariant spinors and tensors can be obtained by decomposing the
original spinors and tensors that form a vector space, or module, for
a given representation of $O(d)$ or $Spin(d)$.  If there are spinors
and tensors of $O(d)$ or $Spin(d)$ admitting invariant components
under $G$, the corresponding vector bundle must be trivial, and thus
it will admit a globally defined non--vanishing section, $i.e.$,
$G$--invariant spinors and tensors.  A nice representation for these
$G$ invariant tensors are the differential forms constructed as
bi-linears of the internal Killing spinors (i.e., the $G$-invariant
spinors)
\be019
\theta_i \gamma_{a_1 \cdots a_n} \theta_j \ .
\ee
The group $G$ is fixed by the number of independent spinors $\theta_i$
which are all singlets under $G$.  E.g.\ if there is only a single
spinor on the 7-manifold, it can be chosen as a real $G_2$ singlet; if
there are two spinors, one can combine them into a complex $SU(3)$
singlet and four spinors as $SU(2)$ singlets.  Of course, all eight
spinors cannot be a singlet of a non-trivial subgroup of $Spin(7)$ and
$G$ is trivial. The 7-dimensional $\gamma$-matrices are in the
Majorana representation and satisfy the relation: $(\gamma_{a_1 \cdots
a_n})^T =(-)^{n^2 +n \over 2}\gamma_{a_1 \cdots a_n}$, which implies
that the differential forms (\ref{019}) are antisymmetric in $[i,j]$
if $n=1,2,5,6$ and otherwise symmetric [we assumed here of course that
$\theta^i$ are commuting spinors and the external spinors are
anti-commuting].

Using complex notation, we can introduce the following two sets of
bi-linears:
\[
\Sigma_{a_1 \cdots a_k } \, \equiv \,
     {\theta}^{\dagger}\gamma_{a_1 \cdots a_k} {\theta}
\qquad \mbox{and}\qquad
\Omega_{a_1 \cdots a_k} \, \equiv \,
     {\theta}^T\gamma_{a_1 \cdots a_k} {\theta}
\]
where  we have suppressed  indices  $i,j$ which count the spinors. The
associated $k$-forms become
\be624
\Sigma^k \, \equiv \,
     {1 \over k!} \Sigma_{a_1 \cdots a_k} e^{a_1} \wedge
 \cdots \wedge e^{a_k}
\qquad \mbox{and}\qquad
\Omega^k \, \equiv \,
     {1 \over k!} \Omega_{a_1 \cdots a_k}  e^{a_1} \wedge
 \cdots \wedge e^{a_k}
\ee
with $e^{a_i}$ as Vielbein 1-forms.  If the spinors are covariantly constant
(with respect to the Levi-Civita connection) the group $G$ coincides with the
holonomy of the manifold. If the spinors are not covariantly constant, neither
can be these differential forms and a deviation of $G$ from the holonomy group
is measured by the intrinsic torsion.  In the following we will discuss
different cases in more detail.

The existence of a $G$-structure lifts the Levi-Civita connection
$\nabla$ to a generalized connection $\nabla^{(T)}$ and the intrinsic
torsion is $\nabla^{(T)}-\nabla$, which can be decomposed into
$G$--modules and has values in $\Lambda^1\otimes \Lambda^2$ (where
$\Lambda^{(k)}$ is space of $k$-forms). Since $\Lambda^2\cong
\mathfrak{so}(d)=\mathfrak{g}\oplus \mathfrak{g}^\perp$ where
$\mathfrak{g}^\perp$ is the orthogonal complement of the Lie algebra
$\mathfrak{g}$ in $\mathfrak{so}(d)$, we conclude that
$(\nabla-\nabla^{(T)})$ can be identified with an element $\tau$ of
$\Lambda^1\otimes {\mathfrak g}^\perp$.  Then the $G$-structure will
be specified by which of these modules, $i.e$, torsion classes, are
present.

On the other hand, the supersymmetric Killing spinor equations in supergravity
theories demand the existence spinors which are parallel with respect to a
generalized connection comprises the Levi--Civita connection as well as the
fluxes contributions
\begin{eqnarray}
\nabla^{(T')}\eta = 0\ .
\label{susyreq}
\end{eqnarray}
As a result, we can rewrite all flux terms in the Killing spinor equations
as
\begin{eqnarray}
\nabla_a^{(T')}\theta\equiv(\nabla-\frac{1}{4}\tau_a^{bc}\gamma_{bc})\theta=0
\end{eqnarray}
and then study supersymmetric solutions and the deformed geometry by analyzing
its group structure in terms of the intrinsic torsion. After identifying the
non-zero torsion components, one can consult the mathematical literature where
examples of these space have been discussed, e.g.\ \cite{630}.  It is
therefore very important to express supersymmetry conditions as constraints on
the intrinsic torsion, and at the same time to classify the possible group
structures in terms of the irreducible components of the intrinsic torsion.


\subsection{$G_2$-Structures}


On 7-dimensional spin manifold $X_7$, $Spin(7)$ is the maximal
structure group with $G_2$ as the maximal subgroup.  Under $G_2$, the
representations of $Spin(7)$ are decomposed as
\begin{eqnarray}
{\rm spinor}\ : \ \ \  {\bf 8}  &\to& {\bf 1}+{\bf 7} \nonumber \\
\Lambda^1\ : \ \ \ {\bf 7}  &\to& {\bf 7} \nonumber \\
\Lambda^2\ : \  {\bf 21} &\to& {\bf 7}+ {\bf 14} \nonumber \\
\Lambda^3\ : \  {\bf 35} &\to& {\bf 1}+ {\bf 7}+{\bf 27}
\label{40}
\end{eqnarray}
The two singlets are the $G_2$ invariant spinor and the $G_2$ invariant rank
three antisymmetric tensor, which can be represented as a bi-linear expression
of the singlet spinor.  The decomposition of the space of 2-forms in
irreducible $G_2$-modules is
\begin{eqnarray}
\Lambda^2 \, = \, \Lambda^2_7 \, \oplus \, \Lambda^2_{14} \ ,
\label{988}
\end{eqnarray}
where
\[
\ba{rcl} \Lambda^2_7 & = & \{u\haken\varphi | u\in TX_7 \}
    =  \{ \alpha \in \Lambda^2 \, |
           \ast(\varphi \wedge \alpha)-2\alpha=0 \} \ , \\[2mm]
\Lambda^2_{14} & = & \{ \alpha \in \Lambda^2 \, |
           \ast(\varphi \wedge \alpha)+\alpha=0 \} \cong \mathfrak{g}_2
\ea
\]
with $\varphi$ denoting the $G_2$-invariant 3-index tensor and the definition
of ``$\haken$'' can be  found in the appendix. Therefore  the  operator
$\ast(\varphi \wedge \alpha)$ splits the 2-forms into corresponding
eigenvalues $2$ and $-1$. The   projections $\mathcal{P}_{\ell}$ onto the
$\ell$-dimensional spaces  then read
\bea930
\mathcal{P}_7(\alpha) & = & {1 \over 3} \,
              (\alpha + \ast(\varphi \wedge \alpha))
    =  {1 \over 3} \, ( \alpha + {1 \over 2} \alpha\haken\psi) \ ,\\
\mathcal{P}_{14}(\alpha)&  = & {1 \over 3} \,
              (2\alpha - \ast(\varphi \wedge \alpha))
   =  {2 \over 3} \, ( \alpha - {1 \over 4} \alpha\haken\psi)
\eea
where $\psi = \ast \varphi$.  To be concrete, the $G_2$-singlet spinor
$\theta_0$ satisfies the condition
\[
({\cal P}_{14})_{ \ ab}^{cd}\, \gamma_{cd} \, \theta_0 = {2 \over
3} \Big( \I^{cd}_{\ \ ab} - {1 \over 4} \psi^{cd}_{\ \ ab} \Big)
\gamma_{cd} \, \theta_0 =0
\]
which is equivalent to
\be552
\gamma_{ab} \theta_0 = i \varphi_{abc} \gamma^c \theta_0 \ .
\ee
Since it is a normalized spinor and due to the properties of 7-d
$\gamma$-matrices (yielding, in particular,   $\theta_0^T \gamma_a \theta_0
=0$), one gets only the following non-zero bi-linears
\begin{eqnarray}
\ba{rcl}
\theta_0^T \theta_0 &=& 1  \ , \\
 \theta_0^T \gamma_{abc} \theta_0 &=& i\, \varphi_{abc} \ , \\
\theta_0^T \gamma_{abcd} \theta_0 &=& - \psi_{abcd}    \ , \\
 \theta_0^T\gamma_{abcdmnp} \theta_0 &=&
i\, \epsilon_{abcdmnp} \ .
\ea
\end{eqnarray}
As we discussed before, $G$-structures can be classified by torsion
classes, which decompose as
\begin{eqnarray}
\tau \ \to \ 7 \times 7 ={\bf 1 + 7 + 14 +27} = \tau_1 + \tau_7 + \tau_{14}
+ \tau_{27}
\end{eqnarray}
with
\be972
\ba{rcl}
&
\ba{rcl}
\tau^{(1)} &\longleftrightarrow&  \psi \haken d\varphi \quad , \\
\tau^{(14)} &\longleftrightarrow&
              \ast d\psi - {1 \over 4} (\ast d\psi) \haken\psi \ ,
\ea
& \qquad
\ba{rcl}
\tau^{(7)} &\longleftrightarrow&  \varphi \haken d\varphi \ , \\
\tau^{(27)} &\longleftrightarrow&
              (d\varphi_{cde\{a}\psi_{b\}}{}^{cde})_0 \ ,
\ea \ea \ee where $\tau_{14}$ and $\tau_{27}$ have to satisfy:
$\varphi \wedge \Lambda_{27}^3= \varphi \wedge \tau_{14} =0$.
Since the Killing spinors define $\varphi$ and $\psi$, these
torsion classes can be obtained from $d \varphi$ and $d \psi$ as
follows
\be093
\ba{rcl}
d\varphi & \in & \Lambda^4 \, = \, \Lambda^4_1 \, \oplus \, \Lambda^4_7
\, \oplus \, \Lambda^4_{27}  \ , \\[2mm]
d \psi & \in & \, \Lambda^5 \, = \, \Lambda^5_7 \, \oplus \,
\Lambda^5_{14}  \ ,
\ea
\ee
where we used (\ref{40}) and the ${\bf 7}$ in $\Lambda^4_7$ is the same as in
$\Lambda^5_7$ up to a constant multiple.


\subsection{$SU(3)$ Structures}


The decomposition of $Spin(7)$ to $SU(3)$ gives
\begin{eqnarray}
{\rm spinor}\ :\  \ \  {\bf 8}  &\to&
{\bf 1}+\bar{\bf 1} + {\bf 3}+\bar{\bf 3} \nonumber \\
\Lambda^1\ : \ \ \  {\bf 7}  &\to& {\bf 1}+{\bf 3} + \bar{\bf 3} \nonumber \\
\Lambda^2\ : \ {\bf 21} &\to& {\bf 1}+2 \times {\bf 3}+2 \times
\bar{\bf 3}+{\bf
8} \nonumber \\
\Lambda^3\ : \ {\bf 35} &\to& {\bf 1}+\bar{\bf 1} + {\bf 1}+ 2
\times {\bf 3}+2 \times \bar{\bf 3}+{\bf 6}+\bar{\bf 6}+{\bf 8}
\label{SU(3) structures}
\end{eqnarray}
We see that $SU(3)$-structures contain two $SU(3)$ invariant real spinors, one
invariant vector field $v$, one invariant 2-form $J$ and a pair of $SU(3)$
invariant three forms $\Psi$, $\bar \Psi$. In total, there are three singlet
3-forms in the decomposition of {\bf 35}, where the extra one corresponds to
$v \wedge J$ and hence it is not independent.

To construct bilinear spinor representations, one
combines the two real singlet spinors into a complex spinor as
\begin{eqnarray}
\theta = {1 \over \sqrt{2}}  ( \I + v_a \gamma^a ) \theta_0 \ ,
\end{eqnarray}
where the constant spinor $\theta_0$ is again the $G_2$ singlet.  The globally
well-defined vector $v$, satisfying $v_a v^a = 1$, gives a foliation of $X_7$
by a 6-manifold $X_6$ and both spinors, $\theta$ and its complex conjugate
$\theta^*$, are chiral spinors on $X_6$. In addition,  there exists a
topological reduction from $G_2$-structures to $SU(3)$-structures (even to
$SU(2)$-structures) and with the vector $v$, an explicit embedding of the
given $SU(3)$-structures in $G_2$-structures reads
\be920
\ba{rcl} \varphi & = & \Psi_+ + v\wedge J \ , \\ \psi & = &
\Psi_- \wedge v \, + \, {1 \over 2} J^2 \ \ea
\ee
where  one defines
\be611
\Psi=\Psi_+ +i \,
\Psi_-
\ee
with $\Psi_-=J.\Psi_+$. Now, the forms, as defined in (\ref{624}), become
\cite{0303127,0311119}
\be910 \ba{rcl} && \Sigma^0  =  1 \ ,  \ \ \ \ \ \ \ \   \Sigma^3
= i v \wedge J \ ,  \nonumber \\&& \Sigma^1  =  v \ , \ \ \ \ \ \
\ \ \Sigma^4 = - {1 \over 2} J \wedge J\ ,\nonumber
\\&& \Sigma^2  =   i  J , \ \ \ \ \ \ \ \ \Sigma^5=-\frac{1}{2} v
\wedge J \wedge J
\nonumber\\&& \Omega^3  =   i \,  \Psi \ , \ \ \ \ \ \  \Omega^4
= -i v \wedge \,  \Psi \ . \ea \ee
and all others vanish. They have to obey the following compatibility
relations
\begin{eqnarray}
\Psi \wedge J =0, && \Psi \wedge \bar \Psi = -\frac{4 i} {3} J
\wedge J \wedge J, \nonumber \\ v \haken J = 0, &&  v  \haken \Psi
=0  \label{compat}
\end{eqnarray}
which follow from the properties of gamma matrices and rearrangements using
Fierz identities. The 2-form, associated with the almost complex structure on
$X_6$ is $J$ and  thus, with the projectors ${1 \over 2} ( 1 \pm i J)$ we can
introduce (anti) holomorphic indices\footnote{Since the 6-d space is in
general not a complex manifold, we cannot introduce global holomorphic
quantities and this projection is justified only locally.}, so that $\Psi$ can
be identified as the holomorphic $(3,0)$-form on $X_6$.

Finally, we present  the torsion classes of $SU(3)$-structures. In the
irreducible $SU(3)$-modules, the $Spin(7)$ 2-forms $\Lambda^2$ decomposes as
\begin{eqnarray}\label{981}
\Lambda^2 = {\bf 21} &\to& {\bf 1}+ 2 \times {\bf 3}+2 \times
\bar{\bf 3}+{\bf 8} \nonumber
\end{eqnarray}
Because the  $SU(3)$ algebra $\mathfrak{g}_{SU(3)} \cong {\bf8}$, the
torsion can be decomposed into
\begin{eqnarray}
\tau & = &({\bf 1}+{\bf 3}+\bar{\bf 3}) \otimes ({\bf 1}+2 \times
{\bf 3}+
2 \times \bar{\bf 3}) \nonumber \\
&\to& \ \  5 \times{\bf 1} \ \ + \ \ 4\times {\bf 8} \ + \ 2\times
({\bf 6} + \bar {\bf
6}) +  5 \times({\bf 3} + \bar {\bf 3})\,, \nonumber \\
&&  ({\cal W}_1+ R+E) + ({\cal W}_2 + T_{1,2})+  ({\cal W}_3 +S) +
({\cal W}_{4,5} +  V_{1,2}+ W_0). \label{eqclass}
\end{eqnarray}
Also in this case, the different components can be read from the
exterior differentials of the forms defining the structure
\begin{eqnarray}
dv &=& R \,
 J + \bar{V}_1 \haken \; \Psi +
V_1 \haken\;\bar{\Psi} + T_1  + v \wedge W_o\,,
 \label{191}\\
d J &=& \frac{3 i}{4}\left(\bar {\cal W}_1\, \Psi -  {\cal W}_1\,
\bar{\Psi}\right) + {\cal W}_3 + J\wedge {\cal W}_4
\nonumber \\
&+& v \wedge \left[\frac13 (E + \bar{E})  J + \bar{V}_2 \haken\;
\Psi + V_2 \haken\; \bar{\Psi} + T_2\right] \,,
\label{192}\\
d\Psi &=& {\cal W}_1  J\wedge J +  J\wedge {\cal W}_2 + \Psi\wedge
{\cal W}_5 + v \wedge \left(E \Psi -4 J \wedge V_2 + S\right)
 \label{193}
\end{eqnarray}
where the numerical coefficients are fixed by compatibility conditions given
by eq.\ (\ref{compat}). Note, the subset ${\cal W}_i$ with $i=1,...,5$ are the
$SU(3)$-structures on the embedded 6-manifold, whose values fix the geometry
of the 6-d base-space \cite{m0202282,0211102,0211118}.


\subsection{$SU(2)$ Structures}


Finally, the $SU(2)$ structures can be obtained by further decomposing
SU(3) representations under $SU(2)$, which yields
\begin{eqnarray}
{\rm spinor} \ : \ \ \ {\bf 8}  &\to& {\bf 1}+\bar{\bf 1} +
({\bf 1} + {\bf 2})+(\bar{\bf 1} + \bar{\bf 2}) \nonumber \\
\Lambda^1 \ : \ \ \ {\bf 7}  &\to& {\bf 1}+({\bf 1} + {\bf 2}) +
(\bar{\bf 1} + \bar{\bf 2}) \nonumber \\
\Lambda^2 \ : \ {\bf 21} &\to& {\bf 1}+2 \times ({\bf 1} + {\bf
2})+ 2 \times (\bar{\bf 1} + \bar{\bf 2})+({\bf
1}+{\bf 2}+{\bf 2} +{\bf 3}) \nonumber \\
\Lambda^3 \ : \ {\bf 35} &\to& {\bf 1}+\bar{\bf 1} + {\bf 1}+ 2
\times ({\bf 1} + {\bf 2})+2 \times (\bar{\bf 1} +
\bar{\bf 2}) \nonumber \\
 && +({\bf
1}+{\bf 2} +{\bf 3})+(\bar{\bf 1}+\bar{\bf 2} +\bar{\bf 3})+({\bf
1}+{\bf 2}+{\bf 2} +{\bf 3})
\end{eqnarray}
There are the following singlets: two complex spinors (or equivalently four
real singlets), one real vector $v^3$ and one complex vector $u=v^1+iv^2$ (or
equivalently three singlet real vectors $v^{\alpha}$ with $\alpha=1,2,3$.),
and one real 2-forms $\omega=\omega^{3}$ and one complex 2-form
$\lambda=\omega^1+i\omega^2$ (or equivalently  three real 2-forms
$\omega^{\alpha}$). These are the basic independent forms. Using these forms
one can build additional three 2-forms and ten 3-forms: $v^{\alpha}\wedge
v^{\beta}$ and $v^{\alpha}\wedge v^{\beta}\wedge v^{\gamma}$,
$v^{\alpha}\wedge \omega^{\beta}$ with $\alpha, \beta = 1, 2, 3$. Note, on any
7-d spin manifold there exist three global vector fields and hence one can
always define $SU(2)$-structures without any assumptions about the manifold.

The $SU(2)$ structures can also be understood by embedding them into
$G_2$, where the three vectors $v_\alpha$ can be chosen as
\[
v_1 \, = \, e^1 \qquad v_2\, = \, e^2  \qquad v_3 \, = \, \varphi(v_1,v_2)
\]
and they parameterize a fibration over a 4-d base space $X_4$. The
embedding of the $SU(2)$ into the $G_2$ structures is then given by
\bea990
\varphi &=& v_1\wedge v_2 \wedge v_3 + v_\alpha\wedge\omega_\alpha \ , \\
\psi &=& vol_4 + \epsilon^{\alpha\beta\gamma} v_\alpha\wedge v_\beta
\wedge \omega_\gamma \ .
\eea
Since the vectors are nowhere vanishing, we can choose them to be of  a unit
norm and perpendicular to each other, i.e.\ $(v_\alpha , v_\beta) =
\delta_{\alpha\beta}$, and using the 3-form $\varphi$, one obtains a
cross-product of these vectors.  One can  choose one of these vectors, say
$v_3$, to define a foliation by a 6-manifold and on this 6-manifold one can
introduce an almost complex structure by $J = v_3 \haken \varphi \ \in \
T^{\ast}M^6\otimes TM^6$.  The remaining two vectors, $u$ and $\bar u$ imply
that this 6-d manifold is a fibration over the 4-d base manifold $X_4$. Note,
on any general 4-d manifold we have the splitting
\begin{eqnarray}
\Lambda^2 = \Lambda^2_+ \oplus \Lambda^2_-
\end{eqnarray}
where $\Lambda^2_+$ and $\Lambda^2_-$ are the selfdual and anti-selfdual
2-forms. We can take $\{\omega_1,\omega_2,\omega_3\}$ as a basis of
$\Lambda^2_-$, which are SU(2)-singlets and satisfy the algebraic relations
\begin{eqnarray}
\omega_i^2 = 2 \, vol_4 \qquad
               \omega_i\wedge\omega_j =0 \quad\mbox{for} \, i\neq j
\end{eqnarray}
and the associating complex structures fulfill the quaternionic
algebra (note: the orientation on the 4-fold is negative). By
further decomposing sub-bundle $\Lambda^2_-$ as
\begin{eqnarray}
\Lambda^2_- \cong  \lambda^{2,0}
         \oplus {\mathbb R} \, \omega
\end{eqnarray}
we can define symplectic structures on this 4-d manifold. In addition to  the
symplectic form $\omega$, the remaining two forms are combined as:
\begin{eqnarray}
\lambda = \omega_1 + i\, \omega_2 \ .
\end{eqnarray}
If the matrix multiplication is denoted by
\begin{eqnarray}
\omega.\lambda \equiv \omega_p^q \lambda_{qr} \label{137}
\end{eqnarray}
the quaternionic algebra\footnote{This algebra can be written in
terms of real components as: $\omega^i . \omega^j = - \delta^{ij} +
\epsilon^{ijk} \omega^k$ .}  implies, that $\lambda$ obeys
\begin{eqnarray}
        (\I - i \omega)_p^q \lambda_{qr} = 0 \label{138}
\end{eqnarray}
where "$\I$" is the identity matrix and hence $\lambda$ is a holomorphic
(2,0)-form. Thus, we recover the three 2-forms, which are $SU(3)$ invariant,
and hence the whole set of $SU(2)$ structures in $G_2$ backgrounds.

A concrete representation of the $SU(2)$ singlet spinors can be given
by
\be992
\theta_1 = \frac{1}{\sqrt{2}} (1 + v_3) \theta_0
\quad , \qquad
\theta_2 = v_1 \, \theta_1
\ee
where $v_\alpha \equiv v^m_\alpha \gamma_m$ and $\theta_0$ as a real
$G_2$ singlet spinor.  Because $(v_1 v_2 -i v_3) \theta_0 = 0$ one
finds
\[
(v_1 - i v_2) \theta_2 = (v_1 + i v_2) \theta_1 = 0 \qquad {\rm or:}
\quad
v_\alpha (\sigma^\alpha)_k{}^l \theta_l = \theta_k \  .
\]
Moreover,
\be620
 \ba{rcl}
v_\alpha v_\beta \theta_k &=& \delta_{\alpha\beta} \theta_k
+ i \epsilon_{\alpha\beta\lambda}
(\sigma^\lambda)_k{}^l \theta_l \ , \\[2mm]
\hat \omega \, \theta_k &=& 4 i \, \theta_k  \ , \\[2mm]
\hat \lambda \, \theta_k  &=&  8 \, (\sigma_2)_k{}^l \theta_l^*
\ea \ee where $\hat \omega \equiv \omega_{mn} \gamma^{mn}$, $\hat
\lambda \equiv \lambda_{mn} \gamma^{mn}$ and with the Pauli
matrices
\be200
\sigma_1=\left( \begin{array}{cc} 0 & 1 \\ 1 & 0 \end{array} \right)
\ ,  \qquad
\sigma_2=\left( \begin{array}{cc} 0 & -i \\ i & 0 \end{array} \right)
\ , \qquad
\sigma_3=\left( \begin{array}{cc} 1 & 0 \\ 0 & -1 \end{array} \right)
\ .
\ee

Based on the embedding above, the $SU(2)$ structures can again be
represented by bi-linears of the complex spinors
\begin{eqnarray}
\ba{ll}
\Sigma^{(0)} = \I \ , & \Omega^{(0)} = 0\ ,\\
\Sigma^{(1)} = v^\alpha \sigma_\alpha \ ,
& \Omega^{(1)} = 0 \ ,\\
\Sigma^{(2)} = i \, \omega \, \I + \Sigma^{(1)} \wedge
\Sigma^{(1)} \ , \qquad
& \Omega^{(2)} = -  \lambda^* \, \sigma_2 \ , \\
\Sigma^{(3)} =  \Sigma^{(1)} \wedge \Sigma^{(2)}  \ ,
& \Omega^{(3)} = - \Sigma^{(1)} \wedge \Omega^{(2)} \ , \\
\Sigma^{(4)} = i\, \Sigma^{(1)} \wedge \Sigma^{(1)}
 \wedge \omega - vol_4 \, \I \ , \
& \Omega^{(4)} = - \Sigma^{(1)} \wedge \Omega^{(3)}  \ , \\
\Sigma^{(5)} =  \Sigma^{(1)} \wedge \Sigma^{(4)}  \ , &
\Omega^{(5)} = \Sigma^{(1)} \wedge \Sigma^{(1)} \wedge
\Sigma^{(1)} \wedge \lambda \sigma_2 . \ea \label{SU(2) bilinears}
\end{eqnarray}
However,  now each form is a $2 \times 2$ matrix. The compatibility of these
forms now imposes
\begin{eqnarray}
\lambda\wedge \lambda =0, \quad \omega^3 \wedge \lambda =0, \quad
\lambda \wedge \lambda^* = 2 \,\omega^3\wedge \omega^3 \,,
\label{SU(2) compatibility}
\end{eqnarray}
as well as
\begin{eqnarray}
u \haken\; \lambda =\bar{u} \haken\; \lambda = 0, \quad u \haken\;
\omega^3 =\bar{u} \haken\; \omega^3 = 0, \quad v^3\haken
\omega^{\alpha}=0 \ .
\end{eqnarray}

As for the torsion classes in $SU(2)$-structures, one can repeat the procedure
done before. The $Spin(7)$ 2-form, $\Lambda^2$, decomposes in the irreducible
$SU(2)$-modules:
\begin{eqnarray}\label{982}
\Lambda^2 = {\bf 21} &\to& {\bf 1}+2 \times ({\bf 1} + {\bf 2})+ 2
\times (\bar{\bf 1} + \bar{\bf 2})+({\bf 1}+{\bf 2}+{\bf 2} +{\bf
3})
\end{eqnarray}
where now the $SU(2)$ algebra is
\begin{eqnarray}
\mathfrak{g}_{SU(2)} = {\bf3}
\end{eqnarray}
and the torsion can be decomposed into a total of 90 classes: 30 singlets, 15
doublets and their complex conjugate and 30 triplets. We will refrain from
presenting  a detailed discussion of  all the classes.

Finally, we need to point out that there is one class of special
$SU(2)$-structures.  Namely, if there are only three real internal
Killing spinors instead of four.  They are all $SU(2)$ singlet
spinors and we shall find this special $SU(2)$-structures in the
\N = 3 and \N = 2 cases below.


\section{Warped Compactification in the Presence of Fluxes}
\resetcounter

\subsection{Supersymmetry Variations}


Compactifications of M-theory in the presence of 4-form fluxes imply in the
generic situation not only a non-trivial warping, but yield a 4-d space time
that is not flat, anymore. This is a consequence of the fact, that for generic
supersymmetric compactifications, the fluxes generate masses for the
gravitinos, which in the simplest case  is proportional to the superpotential.
Hence, a (negative) cosmological constant is generated and the external space
cannot be flat, but becomes an anti-de Sitter (AdS) vacuum. Note, we are not
interested in the situation, where the 4-d superpotential exhibits a run-away
behavior resulting in a singular external space. We consider therefore only
the Ans\" atze for the metric and the 4-form field strength are of the form
\begin{eqnarray}\label{012}
 ds^2 &=& e^{2 U} \Big[ \, g^{(4)}_{\mu\nu} dx^\mu dx^\nu
        + \, h_{ab} dy^a dy^b \Big] \ , \\
\mathbb F &=&{m \over 4!} \, \epsilon_{\mu\nu\rho\lambda} dx^\mu
\wedge dx^\nu \wedge
    dx^\rho \wedge dx^\lambda +
{1 \over 4!} F_{abcd} \, dy^a \wedge dy^b \wedge dy^c \wedge dy^d
\end{eqnarray}
where the warp factor $U=U(y)$ is a function of the coordinates of the
7-manifold with the metric $h_{ab}$, and the 4-d metric $g^{(4)}_{\mu\nu}$ is
either flat or AdS. The Freund-Rubin parameter $m$ corresponds to an unique
flux component along the external space-time which does not violate the 4d
Poincar\'e invariance\footnote{In this paper, we take the convention that
$\epsilon$ denotes an antisymmetric tensor with respect to $\{g_{\mu\nu}, \
h_{ab}, \ G_{AB} \}$ and $\varepsilon$ denotes the associated tensor density.
That is, $\epsilon_{\mu\nu\lambda\rho}=
\sqrt{-g}\varepsilon_{\mu\nu\lambda\rho}$,
$\epsilon_{abcdefg}=\sqrt{h}\varepsilon_{abcdefg}$ and
$\epsilon_{p_0p_1...p_{10}}=\sqrt{-G_{11}}\varepsilon_{p_0p_1...p_{10}}$.}.

Unbroken supersymmetry requires the existence of a Killing
spinor $\eta$ yielding a vanishing gravitino variation of
11-dimensional supergravity
\begin{eqnarray}
0 = \delta \Psi_M =
        \Big[ D_M
        + {1 \over 144} \Big(\Gamma_M^{\ NPQR} - 8 \, G_M^N\,
        \Gamma^{PQR} \Big) \, \mathbb F_{NPQR} \Big] \eta
\end{eqnarray}
with $D_M=\partial_M + {1 \over 4} \hat \omega^{RS}_M \Gamma_{RS}$. In
a first step one transforms from the warped or conformal frame to
the non-warped or original frame. Note, this transformation is not
a change of coordinates, but an actual change of geometry.  Using
\begin{eqnarray}
ds^2 = e^{2U} \widetilde{ds}^2 \quad \rightarrow \quad D_M =
\tilde D_M + {1 \over 2}    \Gamma_M^{\ N} \partial_N U
\end{eqnarray}
we have
\begin{eqnarray}
0 &=& \Big[ \tilde D_M +\frac{1}{2} \Gamma_M^{\ N} \partial_N U +
        {e^{-3U} \over 144} \Big(\Gamma_M^{\ NPQR} - 8 \, G_M^N\,
        \Gamma^{PQR} \Big) \, \mathbb F_{NPQR} \Big] \eta \nonumber \\
        &=& \Big[ \tilde D_M +\frac{1}{2} \Gamma_M^{\ N} \partial_N U
        + {e^{-3U} \over 144} \Big(\Gamma_M \hat \mathbb F - 12 \,
        \hat \mathbb F_M \Big) \Big] \eta \label{10}
\end{eqnarray}
with all $\Gamma$ matrices defined in the original frame and all
indices raised and lowered in the original frame. We used here
identities for $\Gamma$-matrices (see appendix) and introduced the
abbreviation
\begin{eqnarray}
\hat \mathbb F \equiv \mathbb F_{MNPQ} \Gamma^{MNPQ} \quad, \qquad
\hat \mathbb F_M \equiv \mathbb F_{MNPQ} \Gamma^{NPQ} \label{112}
\end{eqnarray}
where $\mathbb F_{MNPQ}$ is the same as that in the conformal frame.
All indices in eq.\ (\ref{10}) are curved, but most of the calculation
is done in the tangent space.  Only for the discussion of the
$G$-structure differential equations, we have to go back to the original
coordinates.

Similar to the metric and 4-form, we  also have to split the spinor into an
external and internal spinor and with the notation from the appendix, the flux
is decomposed as
\begin{eqnarray}
 \hat \mathbb F = - i  \, m \, \hat \gamma^5 \otimes {\mathbb{I}} +
\mathbb{I}\otimes F\ ,
\\
\hat \mathbb F_\mu = {i \over 4} \, \, m  \hat \gamma^5 \hat
\gamma_\mu \otimes
   {\mathbb I} \quad , \qquad \hat \mathbb F_a = \hat \gamma^5 \otimes F_a
\end{eqnarray}
where $F$ and $F_a$ are defined as in (\ref{112}), but using the 7-d
$\gamma^a$-matrices instead of the 11-d matrices. The gravitino
variation splits therewith into an external and internal part
\begin{eqnarray}
0&=&\Big[ \tilde\nabla_\mu \otimes \mathbb{I} +  \hat \gamma_\mu
\hat \gamma^5
  \otimes \Big( {1 \over 2} \, \partial U + {i \, m \over 72}e^{-3U} \Big)
+ {1 \over 144} e^{-3U} \, \hat \gamma_\mu \otimes F
  \Big] \eta \label{109} \ , \\
0 &=& \Big[ \mathbb{I} \otimes \Big( \nabla_a  - {i\, m \over 144}
e^{-3U}\, \gamma_a \Big) +
        {1 \over 144} e^{-3U}\,  \hat \gamma^5 \otimes
  \Big( \gamma_a F  -12 F_a \Big) \Big] \eta  \label{110}
\end{eqnarray}
here $\partial U \equiv \gamma^a \partial_a U$, and
$\nabla_a=\tilde\nabla_a + {1 \over 2} \gamma_a^{\ b} \, \partial_b
U$. The different covariant derivatives $\{ \tilde\nabla_\mu ,
\tilde\nabla_a , \nabla_a\}$ are related to the metrics $\{g_{\mu\nu},
\ h_{ab}, \ e^{2U}h_{ab}\}$.

In eq.\ (\ref{110}), we can eliminate the term $\sim \gamma_a F \eta$
by multiplying eq.\ (\ref{109}) with (${1 \over 4} \hat \gamma^5 \hat
\gamma^\mu\otimes \gamma^a$) and subtracting both expressions.
We find
\begin{eqnarray}
0 = \tilde \eta + \Big[\hat \gamma^5 \otimes \Big( {1 \over 2}
\partial U + {i m \over 72} e^{-3U}\Big) +{1 \over 144} e^{-3 U}\,
(\mathbb I \otimes F) \Big] \eta \ , \label{external KQ0} \\
0=
\mathbb{I} \otimes \Big( \nabla_a -\frac{1}{2} \gamma_a\partial U-
{ i m \over 48} e^{-3U} \gamma_a \Big)\eta - \hat \gamma^5 \otimes
\gamma_a \tilde \eta - {1 \over 12}\, e^{-3U}\, \hat \gamma^5
\otimes F_a  \eta \label{internal KQ0}
\end{eqnarray}
where $\tilde \eta$ is defined by
\be953
\Big[\tilde\nabla_\mu \otimes \mathbb I \Big] \eta = (\hat
\gamma_\mu \otimes \mathbb I ) \tilde \eta  \ .
\ee
This spinor is non-zero whenever the external spinor is not covariantly
constant and below we will give the explicit form.  In fact, due to the fluxes
both the 7- as well as the 4-d spinors are not anymore covariantly constant
and the deviation is measured by torsion classes, which have been discussed in
detail in the previous section. On the external space, this back-reaction of
fluxes is measured by the 4-d cosmological constant and thus the space becomes
AdS.  In a supersymmetric vacuum the cosmological constant is given by the
determinant of the mass matrix of the gravitinos and the corresponding Killing
spinors cannot be covariant constant. Therefore, we  the 4-d spinors solve the
equation
\be811
\tilde\nabla_\mu \epsilon^x \  \sim \ \hat \gamma_\mu \,
 ( W_1^{xy} + i \hat \gamma^5 \,W_2^{xy} )
\, \epsilon_x \
\ee
and the gravitino mass matrix is an element of the $R$-symmetry of the
underlying supergravity.  If there is only one external spinor (\N = 1 case),
this gravitino mass matrix is simply the superpotential and the equation
simplifies:
\[
\tilde\nabla_\mu \epsilon ~ \sim ~ \hat \gamma_\mu  \,
(W_1 + i \, \hat \gamma^5 \, W_2 ) \, \epsilon \ .
\]
If $\epsilon$ is a Weyl spinor, this equation becomes
$\tilde\nabla_\mu \epsilon = \hat \gamma_\mu \bar W \epsilon^*$ with
the complex superpotential
\be252
W = W_1 +i\, W_2 \ .
\ee
More precisely,  we should  also take into account a non-trivial K\"ahler
potential $K$ by replacing $W \rightarrow e^{K/2} W$; only this rescaled
quantity has the proper holomorphicity  structure of  4d \N =1 supergravity.

Let us add a comment on the internal spinor equations. If the flux
contribution in the second term of eq.\ (\ref{110}) (which is proportional to
$\hat \gamma^5$) decouples from the first term in eq.\ (\ref{110}), the
internal space has to be an Einstein space and can be lifted to an 8-d space
of special holonomy.  There are three cases of special interest, which have
been also discussed  in the mathematical literature \cite{FrKa,BoGa,Bar};
these three classes are related to the number of real internal spinors. If
there is a single internal spinor, the 7-d space has $G_2$ structures and can
be lifted to an 8-d space of $Spin(7)$ holonomy; for two real spinor we have
$SU(3)$ structures and the lift yields a space of $SU(4)$ holonomy
(Calabi-Yau); finally the case with three real spinors can be lifted to an 8-d
hyper K\"ahler space. Correspondingly, the 7-d Einstein space is a weak $G_2$,
Einstein-Sasaki or tri-Sasaki manifolds. Note that  the last case is a {\em
special} examples of $SU(2)$ structures, whereas for the general case with
$SU(2)$ structure, ie.\ the case with four real spinors, the corresponding 8-d
space is not simply connected, because in this case, the 8-d space has to have
a covariantly constant vector and the holonomy is $SU(2)$. This is obvious,
because 8-d manifolds that do not factorize, can have only $Spin(7)$, $SU(4)$
and $Sp(2)$ as restricted holonomy.

Before we can discuss  different supersymmetric flux vacua, we have to split
the Killing spinor $\eta$ into internal and external spinors which will be
addressed in the following subsection.


\subsection{Killing Spinors}


Group structures are specified by Killing spinors, which are invariant under
the corresponding structure group, and its embedding into $Spin(7)$ is
parameterized by globally well-defined vectors. As we classified before,
$G_2$-structures admit a singlet spinor on the 7-manifold; $SU(3)$-structures
admit a complex $SU(3)$ singlet; $SU(2)$-structures admit two complex or three
real $SU(2)$ singlet spinors. If there are even more Killing spinors, no
$G$-structure can be defined.  The most general Killing spinor in M-theory,
specifies $SU(2)$-structures and can preserve up to \N = 4 supersymmetry. All
other Killing spinors preserve either a larger structure group or less
supersymmetry and can be obtained by introducing new projectors, which further
reduce the number of 7-d singlet spinors or 4-d external ones. The
corresponding classification of the 11-d Killing spinors are given as follows.

\bigskip

\noindent {\bf (1) \N = 4}

\medskip

\noindent The most general 4-d \N=4 Killing spinor with
non-trivial G-structures\footnote{In this paper we only consider
Killing spinors with non-trivial G-structures, where the first
non-trivial structure group is $SU(2)$. Actually, there are also
\N=4 vacua, where the structure group is trivial and hence
Eq(\ref{N=4}) is not the most general one.} can be obtained by
expanding the 11-d Majorana spinor as
\begin{eqnarray}
\eta &=& \sum_{x=1}^2 (a_{xL}\epsilon_{xL} +
a_{xR}\epsilon_{xR})\otimes \theta_x + cc \ . \label{N=4}
\end{eqnarray}
where $a_{xL} \epsilon_{xL} + a_{xR} \epsilon_{xR}$ denotes 4-d Dirac spinors
with $\epsilon_{R/L}$ as its chiral components and $\theta_x$ are 7-d $SU(2)$
singlet spinors. Note, $\theta_x$ are normalized, which is also true for other
cases. In the following spinor projectors we use the doublet notation

\renewcommand{\arraystretch}{1}

\[ \epsilon_x \equiv \Big( \ba{c} \epsilon_{xL} \\
\epsilon_{xR} \ea \Big) \] \ .

\bigskip

\noindent {\bf (2) \N = 3}

\medskip

\noindent To obtain an \N=3 Killing spinor Ansatz, we need to
truncate the spinor (\ref{N=4}), which can be achieved by the
spinor projector
\begin{eqnarray}
  \epsilon_2=\sigma^1\epsilon_2^* \ , \label{03}
\end{eqnarray}
which leads to the most general 4-d \N=3 Killing spinor
\begin{eqnarray}
  \eta &=& (a_{1L}\epsilon_{1L} + a_{1R}\epsilon_{1R})\otimes
  \theta_1 +(a_{2L}\epsilon_{2L} + a_{2R}\epsilon_{2L}^*)\otimes
  \theta_2 + cc \nonumber \\
&=& (a_{1L}\epsilon_{1L} + a_{1R}\epsilon_{1R})\otimes
\theta_1+\epsilon_{2L}\otimes( a_{2L}\theta_2 +
a_{2R}^*\theta_2^*) + cc \ . \label{N=3(I)}
\end{eqnarray}
Note, $a_{2L}$ and $a_{2R}$, and $a_{1L}$ and $a_{1R}$ cannot be
simultaneously zero, otherwise, eq. (\ref{N=3(I)}) is  reduced to $SU(3)$ \N=2
and \N=1 Killing spinors, respectively. This general \N=3 Killing spinor
Ansatz preserves $SU(2)$-structures, and it has some special cases:

\medskip

$(a)$ $a_{2L}=0$ or $a_{2R}=0$, which gives three 4-d Weyl spinors
\begin{eqnarray}
\eta &=& (a_{1L}\epsilon_{1L} + a_{1R}\epsilon_{1R})\otimes
\theta_1 +
a_{2L}\epsilon_{2L} \otimes \theta_2 +cc \ , \\
 {\rm or}\ \ \ \ \eta &=&
(a_{1L}\epsilon_{1L} + a_{1R}\epsilon_{1R})\otimes \theta_1 +
a_{2R}\epsilon_{2L}^*\otimes \theta_2 +cc
 \label{N=3(I)a}
\end{eqnarray}
or

\medskip

$(b)$ $a_{2L}=a_{2R}^*$ and hence there are two 4-d Weyl and one
Majorana spinor
\begin{eqnarray}
\eta &=& (a_{1L}\epsilon_{1L} + a_{1R}\epsilon_{1R})\otimes
\theta_1 + (a_{2L}\epsilon_{2L} + a_{2L}^* \epsilon_{2L}^*)\otimes
\theta_2 +cc
\nonumber \\
&=& (a_{1L}\epsilon_{1L} + a_{1R}\epsilon_{1R})\otimes \theta_1 +
a_{2L}\epsilon_{2L}\otimes(\theta_2 + \theta_2^*) + cc \ .
 \label{N=3(I)b}
\end{eqnarray}
Note, even though this case preserves $SU(2)$-structures, it is very special
since it has only three real internal Killing spinors, compared to four
spinors for general $SU(2)$ cases.

\bigskip

\noindent {\bf (3) \N = 2 (I)}

\medskip

\noindent There are two ways to truncate the \N = 3 Killing spinor Ansatz: one
preserves  $SU(2)$-structures and the other one preserves $SU(3)$-structures.
The new projector that preserves $SU(2)$-structures reads
\begin{eqnarray}
  \epsilon_1=\sigma^1\epsilon_1^* \label{05}
\end{eqnarray}
which gives the most general $SU(2)$ \N=2 Killing spinor
\begin{eqnarray}
  \eta &=& (a_{1L}\epsilon_{1L} + a_{1R}\epsilon_{1L}^*)\otimes
  \theta_1 +(a_{2L}\epsilon_{2L} + a_{2R}\epsilon_{2L}^*)\otimes
  \theta_2 + cc \nonumber \\
  &=& \epsilon_{1L}\otimes( a_{1L}\theta_1 +
  a_{1R}^*\theta_1^*)+\epsilon_{2L}\otimes( a_{2L}\theta_2 +
  a_{2R}^*\theta_2^*) + cc \ . \label{N=2(I)}
\end{eqnarray}
Similar to \N=3 case, this general Killing spinor has some special cases as
well. Here we list three which we will  discuss in this paper:

\bigskip

$(a)$ If $a_{1R}=a_{2R}=0$ (or  $a_{1L}=a_{2L}=0$)
\begin{eqnarray}
  \eta &=& a_{1L}\epsilon_{1L}\otimes \theta_1 + a_{2L}\epsilon_{2L}
  \otimes \theta_2 +cc
  \\ {\rm or}\ \ \ \  \eta &=& a_{1R}\epsilon_{1L}^*\otimes \theta_1 +
  a_{2R}\epsilon_{2L}^*\otimes \theta_2 +cc
 \label{N=2(I)a}
\end{eqnarray}
For this case we have two 4-d Weyl spinors of the {\em same}
chirality.

\bigskip

$(b)$ $a_{1R}=a_{2L}=0$ (or  $a_{1L}=a_{2R}=0$)
\begin{eqnarray}
  \eta &=& a_{1L}\epsilon_{1L}\otimes \theta_1 +
  a_{2R}\epsilon_{2L}^* \otimes \theta_2 +cc
  \\ {\rm or}\ \ \ \  \eta &=& a_{1R}\epsilon_{1L}^*\otimes \theta_1 +
  a_{2L}\epsilon_{2L}\otimes
\theta_2 +cc
 \label{N=2(I)b}
\end{eqnarray}
For this case we have two 4-d Weyl spinors of {\em opposite}
chirality.

\bigskip

$(c)$ For $a_{2L}=a_{2R}^*$
\begin{eqnarray}
\eta &=& (a_{1L}\epsilon_{1L} + a_{1R}\epsilon_{1L}^*)\otimes
\theta_1 + (a_{2L}\epsilon_{2L} + a_{2L}^*\epsilon_{2L}^*)\otimes
\theta_2 +cc
\nonumber \\
&=& (a_{1L}\epsilon_{1L} + a_{1R}\epsilon_{1L}^*)\otimes \theta_1
+ a_{2L}\epsilon_{2L}\otimes(\theta_2 + \theta_2^*) + cc
 \label{N=2(I)c}
\end{eqnarray}
yielding one 4-d Weyl spinor and one 4-d Majorana spinor. This
case is similar to case $(b)$ of \N = 3: it has only three real
internal Killing spinors.

\bigskip

\noindent {\bf (4) \N = 2 (II)}

\medskip

\noindent The general $SU(3)$ \N = 2 Killing spinor Ansatz can be
obtained by setting $a_{2L}=a_{2R}=0$ in the Ansatz (\ref{N=4}) or
(\ref{N=3(I)}).  The truncated spinor becomes
\begin{eqnarray}
\eta = (a_L\epsilon_L + a_R\epsilon_R)\otimes \theta + cc
\label{N=2a}
\end{eqnarray}
where $\theta = \theta_1$, $a_L=a_{1L}$ and $a_R=a_{1R}$.

\bigskip

\noindent {\bf (5) \N = 1 (I)}

\medskip

\noindent Finally, in order to obtain \N = 1 Killing spinor Ansatz, we have to
take another truncation for \N = 2 Killing spinor (\ref{N=2a}). There are
again two ways: one yielding $SU(3)$- and the other one $G_2$-structures. The
general $SU(3)$ \N=1 Killing spinor is obtained by setting
\begin{eqnarray}
\epsilon=\sigma^1\epsilon^* \label{09}
\end{eqnarray}
which leads to
\begin{eqnarray}
\eta & = & (a_L \epsilon_L + a_R \epsilon_L^*) \otimes \theta + cc
\nonumber \\&=& \epsilon_L \otimes (a_L\theta + a_R^* \theta^*) +
cc  \ . \label{N=1(I)}
\end{eqnarray}
Note, here $a_L\neq a_R^*$, otherwise eq. (\ref{N=1(I)}) is reduced to $G_2$
\N=1 Killing spinor. This general $SU(3)$ \N=1 Killing spinor Ansatz can be
{\it completely} classified by the following two cases: (I) $a_L=0$ or
$a_R=0$; (II) $a_La_R \neq 0$. For the first case, the Killing spinor Ansatz
is
\begin{eqnarray}
\eta & = & a \epsilon \otimes \theta + cc \label{N=1(I)a} \ ,
\end{eqnarray}
and the associated flux solutions have been discussed extensively
in the literature \cite{0303127,0311119, 0311146}. The second
class is a new one which  has not  been discussed before.

One may wonder whether case (II) is equivalent to case (I). This
would be the case if we can rewrite the spinor in case (II) as
\begin{eqnarray}
  a_L \theta + a_R^* \theta^* = b \tilde \theta \label{601}
\end{eqnarray}
where $a_La_R\neq 0$ and $\tilde \theta$ is a new $SU(3)$ singlet
spinor. The necessary condition for this is that one can embed
$\tilde \theta$ in $G_2$-structures
\begin{eqnarray}
  \tilde \theta = (1 + \tilde v) \tilde \theta_0       \label{602}
\end{eqnarray}
with a real spinor $\tilde \theta_0$ and $\tilde v$ being a new
rotated global vector, which specifies the new $SU(3)$-structures.
It would have to satisfy
\begin{eqnarray}
        (1 - \tilde v) \tilde \theta = 0   \ .          \label{603}
\end{eqnarray}
However, due to eq. (\ref{552}), one finds that the reality condition for the
new global vector $\tilde v$ cannot be guaranteed for arbitrary complex $a_L$
and $a_R$, thus  implying that the assumption in eq.\ (\ref{601}) is
incorrect.

Actually, Killing spinor in case (I) has definite 10-d chirality (since
$\theta$ is 6-d chiral spinor) and it can survive $S_1/Z_2$ projection with
$S_1$ being $SU(3)$ fiber in the internal space. So, it can be embedded in the
Horava-Witten scenario of  heterotic M-theory. As for the Killing spinor in
case (II), it only admits $S^1$ fiberation of $SU(3)$ structures and hence it
matches with the M-theory lift of  type IIA supergravity. Since case (II) has
not yet been discussed in the literature, we will explicitly discuss it in
this paper.

\bigskip

\noindent {\bf (6) \N = 1 (II)}

\medskip

\noindent If $a_L=a_R^*$, the spinor (\ref{N=1(I)}) becomes
\begin{eqnarray}
\eta = a \epsilon \otimes \theta \  .   \label{N=1(II)}
\end{eqnarray}
with $a$ real, and $\epsilon$, $\theta$ Majorana spinors. This
spinor Ansatz is the unique one preserving $G_2$-structures.


\subsection{Bianchi Identities and Equations of Motion}


Killing spinor equations give only necessary conditions for supergravity
solutions. For maximally supersymmetric solutions, the equations of motion are
equivalent to the integrability constraints of the Killing spinor equations.
Hence, if certain supersymmetries are broken by the solution, they are not
automatically satisfied (because in this case certain components of the
Killing spinor equations are projected out).  To complete the calculation and
provide sufficient conditions, we need to consider Bianchi identities and
equations of motion as well.  The 11-d source-free Bianchi identity is given
by
\begin{eqnarray}
  (d\mathbb F)_{11}=0 \label{151}
\end{eqnarray}
and can be separated into the external and internal part
\begin{eqnarray}
(d m)_7 =0  \ , \ \ (dF)_7=0 \ . \label{152}
\end{eqnarray}
The external Bianchi identity implies that the Freund-Rubin parameter
is constant. The index ``7'' indicates that the exterior derivative is
taken in the 7-d internal space. The equation of motion becomes
\begin{eqnarray}
  (d * \mathbb F)_{11} = \mathbb F \wedge \mathbb F \label{153}
\end{eqnarray}
and due to
\begin{eqnarray}
  * \mathbb F &=& e^{4U}(* m)_7 + e^{4U}(* F)_7 dV_4\ , \\
\mathbb F \wedge \mathbb F &=& 2 m F dV_4\ .
\end{eqnarray}
Because $(* m)_7$ is proportional to the 7-d volume form and
$U$ depends only on the internal coordinates, we get
\begin{eqnarray}
d[e^{4U}(* F)_7 ] = 2 m F \label{159}
\end{eqnarray}
where $dV_4$ is the volume element of the 4-d (un-warped) external space-time,
which we have cancelled on both sides of the equation.

Let us add additional  comments regarding  Bianchi identities and equations of
motion: (i) the dualization in $({*} F)_7$ is done with respect to the warped
metric giving a contribution $e^{-U}$ from the warp factor; (ii) it may happen
that eq.\ (\ref{151}) and eq.\ (\ref{153}) can only be satisfied if sources
(ie.\ M-branes) are taken into account and we should keep in mind that the
Killing spinor equations provide local conditions  and they may not
distinguish  \`a priori  between background fluxes and fluxes sourced by
M-branes.


\subsection{Decomposition of Flux Components}


{\em (1)} For $SU(3)$ structure group, we project the fluxes onto the
base $X_6$  which gives
\begin{eqnarray}
G=F|_{X_6} \ ,\  \ H= F\haken v \ .
\end{eqnarray}
$G$ and $H$ are forms on $X_6$ that decompose under $SU(3)$
\begin{eqnarray}
&& [G]={\bf 15 =8+3+\bar{3}+1}, \nonumber \\
&& [H]={\bf 20=6+\bar{6}+3+\bar{3}+1+\bar{1}}
\end{eqnarray}
These components have the following holomorphic structure
\begin{eqnarray}
&G:&  {\bf 8+1} \  \leftrightarrow \  \Lambda^{(2,2)} \nonumber \\
&& {\bf 3+\bar{3}} \ \leftrightarrow \
  \Lambda^{(3,1)}+\Lambda^{(1,3)} \nonumber \\
&H:& {\bf  6+\bar{6} + 3 + \bar{3}} \ \leftrightarrow \ \Lambda^{(2,1)}+
 \Lambda^{(1,2)}\nonumber \\
    && {\bf 1+\bar{1}} \ \leftrightarrow \ \Lambda^{(3,0)} + \Lambda^{(0,3)}
\end{eqnarray}
among which the {\bf 1} component of $G$  and the ${\bf 3}+{\bf \bar 3}$
of $H$ are non-primitive. Therefore, we write the 4-form as
\begin{eqnarray}
2 \, e^{-3U}F&=& P (J \wedge J) + J \wedge T + \Psi \wedge \bar V
\nonumber \\
&&+ v \wedge (Q \, \Psi + J \wedge V_o + H^{6+\bar 6}  ) +cc \label{171}\ .
\end{eqnarray}
Here, the real 2-form $T$ denotes the {\bf 8} components, $V$ and
$V_o$ denote ${\bf 3+ \bar 3}$ components from $G$ and $H$ part respectively,
with $V$ being complex and $V_o$ being real, and $H^{6+\bar 6}$
denotes the ${\bf 6}+\bar {\bf 6}$ components of the $H$ part
fluxes. Correspondingly, the coefficients and the associated flux
components can be expressed as
\begin{eqnarray}
P&=&\frac{e^{-3U}}{3!3! 4!} (J \wedge J)\haken F \nonumber \ , \\
Q&=& \frac{e^{-3U}}{4 \times 4!}(v\wedge \bar \Psi) \haken F
\label{172}
\end{eqnarray}
and
\begin{eqnarray}
  T&=& \frac{e^{-3U}}{4} J \haken F -PJ - \frac{5}{4} v \wedge V_o
   \ , \nonumber \\
  \bar V &=& \frac{e^{-3U}}{24} \bar \Psi \haken F +  v \, Q\nonumber \ , \\
  V_o &=& \frac{e^{-3U}}{12} (v \wedge J) \haken F \ , \nonumber \\
  H^{(6+\bar 6)} & = & \frac{e^{-3U}}{4} v \haken F - {1 \over 2}
  (Q \, \Psi + J \wedge V_o + cc) \ .
\end{eqnarray}
For the  Hodge dual of $F$ we introduce the notation
\begin{eqnarray}
{*}(J\wedge J) &=&z_1\sqrt{G_7} \, v \wedge J \nonumber \ , \\
{*} (J \wedge T) &=& z_2 \sqrt{G_7} \, v \wedge T \nonumber\ , \\
{*} (\Psi \wedge \bar V) &=& z_3 \sqrt{G_7} \, v \wedge (V \haken \bar \Psi )
 \nonumber\ , \\
{*} (v \wedge \Psi) &=& z_4\sqrt{ G_7} \, \bar \Psi \nonumber\ , \\
{*} (v \wedge J \wedge V_o) &=&  z_5\sqrt{
G_7}\, J\wedge (V_o \haken J)\nonumber\ , \\
{*} (v \wedge H^{6+\bar 6}) &=&  z_6\sqrt{G_7}\, H^{6+\bar 6}\label{173} \ .
\end{eqnarray}
The associated coefficients $z_i$ can be calculated as follows.  If
one denotes ${*} \xi_i = Z_i \zeta_i$ the coefficients become
\begin{eqnarray}
Z_i=\frac{3! \xi \haken \xi
 }{\varepsilon^{abcdefg}\xi_{abcd} \zeta_{efg}} \label{145} \ .
\end{eqnarray}
Finally, the Hodge dual of $F$ can be written as
\begin{eqnarray}
  2\, \frac{{*} e^{-3U}F}{\sqrt{G_7}}&=&v \wedge (z_1 P \, J + z_2 \,
  T + z_3 \, V \haken \bar \Psi ) \nonumber \\
  &&+ (z_4 Q\, \Psi  + z_5 \, J \wedge (V_o \haken J) +
  z_6 \, H^{6+\bar 6})  + cc\label{174} \ .
\end{eqnarray}

\bigskip
\noindent {\em (2)} If the structure group is $SU(2)$, the 7-d internal space
is written as a 3-d fibration over a 4-d base $X_4$. This fibration is fixed
by the three vectors $v_\alpha$ which  also yield the following projection of
the flux components:
\begin{eqnarray} \label{329}
&&A=F, \nonumber \\ &&B_{\alpha}= F\haken
  v_{\alpha}=F_{abcd}v^a_{\alpha} \nonumber\ , \\
&&C_{\alpha\beta}=F\haken(v_{\alpha}\wedge
  v_{\beta})=2!F_{abcd}v^a_{\alpha}v^b_{\beta} \ ,\nonumber \\
&&D_{\alpha\beta\gamma}=F\haken(v_{\alpha}\wedge v_{\beta}\wedge
  v_{\gamma})=3!F_{abcd}v^a_{\alpha}v^b_{\beta}v^c_{\gamma} \ .
\end{eqnarray}
For future convenience, we also define
\begin{eqnarray}
B=B_{\alpha}\sigma^{\alpha} \ ,\
C=C_{\alpha\beta}\sigma^{\alpha}\sigma^{\beta}\ , \
D=D_{\alpha\beta\gamma}\sigma^{\alpha}\sigma^{\beta}\sigma^{\gamma} \ .
\end{eqnarray}
These base tensors can be decomposed under $SU(2)$, giving
\begin{eqnarray}
&&[A]={\bf 1}\ ,\
[B]={\bf 12}=3\times{\bf (2+\bar{2})}\ , \\
&&[C]={\bf 18}=3\times {\bf (1+3+1+\bar{1})}\ , \ [D]={\bf 4=2+\bar{2}}.
\end{eqnarray}
where we use the symplectic form $\omega$ in (\ref{SU(2) bilinears})
for the holomorphic projection and the factor {\bf 3} appearing in $[B]$ and
$[C]$ are associated with the triplicity of the invariant vector
fields. They can be identified with the following forms on $X_4$
\begin{eqnarray}
 &A:&  {\bf 1}  \leftrightarrow  \Lambda^{(2,2)} \nonumber\ , \\
 &B:& {\bf 2+\bar{2}} \leftrightarrow \Lambda^{(2,1)}+\Lambda^{(1,2)}
  \nonumber\ , \\
 &C:&  {\bf 1+\bar{1}} \leftrightarrow \Lambda^{(2,0)}+\Lambda^{(0,2)}
  \nonumber\ , \\
     && {\bf 1+3} \leftrightarrow \Lambda^{(1,1)} \nonumber\ , \\
 &D:& {\bf  2+\bar{2}} \leftrightarrow \Lambda^{(1,0)}+\Lambda^{(0,1)}\ .
\end{eqnarray}
Obviously, the total number of components of the forms $A$, $B$, $C$ and $D$
match with the components of the 4-form $F$. However,  among these projected
forms only $A$ and $C$ are regular (because the 4-d base space does not
support regular 1-forms or 3-forms). The form $A$ is proportional to the 4-d
volume form and $C$ gives the possible 2-forms on $X_4$.  In the following we
will keep only these regular forms and hence drop $B$ and $D$. Thus, the
4-form flux can be written as
\begin{eqnarray}
e^{-3U}F=X(\omega \wedge \omega) + (v_{\alpha}\wedge v_{\beta})
\wedge (Y^{\alpha\beta}_1\omega + Y^{\alpha\beta}_2\lambda+
Y^{\alpha\beta}_3\lambda^*+ Y_4^{\alpha\beta}) \label{141}
\end{eqnarray}
where the $e^{-3U}$ factor is added here for convenience, because this
combination appears in the Killing spinor equations (\ref{external
KQ0}) and eq.\ (\ref{internal KQ0}). The 2-forms can be expressed as
\begin{eqnarray}
X&=&\frac{e^{-3U}}{2(3! 4!)} (\omega\wedge \omega)\haken F \nonumber \ , \\
Y^{\alpha\beta}_1&=& \frac{e^{-3U}}{2(3! 4!)}(v^{\alpha}\wedge
v^{\beta}\wedge \omega) \haken F  \nonumber \ , \\
Y^{\alpha\beta}_2&=& \frac{e^{-3U}}{4( 3! 4!)}(v^{\alpha}\wedge
v^{\beta}\wedge \lambda^*) \haken F \nonumber \ , \\
Y^{\alpha\beta}_3&=& \frac{e^{-3U}}{4(3! 4!)}(v^{\alpha}\wedge
v^{\beta}\wedge \lambda) \haken F \label{142}
\end{eqnarray}
and
\begin{eqnarray}
Y_4^{\alpha\beta} =\frac{e^{-3U}}{4} (v^\alpha \wedge v^\beta)
\haken F -(Y^{\alpha\beta}_1\omega + Y^{\alpha\beta}_2\lambda+
Y^{\alpha\beta}_3\lambda^*) \ . \label{142a}
\end{eqnarray}
are the {\bf 3} component of $C^{\alpha\beta}$.  For the Hodge dual
components we use
\begin{eqnarray}
* (\omega\wedge \omega)
&=&Z\sqrt{G_7}\varepsilon_{\alpha\beta\gamma} v^{\alpha}\wedge
v^{\beta}\wedge v^{\gamma} \nonumber \ , \\
{*} (v_{\alpha}\wedge v_{\beta}\wedge \omega)&=&
Z_1\sqrt{G_7}\varepsilon_{\alpha\beta\gamma}v^{\gamma}\wedge
\omega
 \nonumber \ , \\
{*} (v_{\alpha}\wedge v_{\beta}\wedge\lambda) &=&
Z_2\sqrt{G_7}\varepsilon_{\alpha\beta\gamma}
v^{\gamma}\wedge\lambda^*  \nonumber \ , \\
{*} (v_{\alpha}\wedge v_{\beta}\wedge\lambda^*) &=&
Z_3\sqrt{G_7}\varepsilon_{\alpha\beta\gamma}
v^{\gamma}\wedge\lambda \nonumber \ , \\
{*} (v_{\alpha}\wedge v_{\beta}\wedge Y_4^{\alpha\beta}) &=&
Z_4\sqrt{G_7}\varepsilon_{\alpha\beta\gamma}v^{\gamma}\wedge
Y_4^{\alpha\beta} \label{143}
\end{eqnarray}
and write
\begin{eqnarray}
\frac{{*} e^{-3U}F}{\sqrt{G_7}}&=&XZ
\varepsilon_{\alpha\beta\gamma} v^{\alpha}\wedge v^{\beta}\wedge
v^{\gamma} \nonumber
\\&& +\varepsilon_{\alpha\beta\gamma}
v^{\gamma} \wedge (Y_1^{\alpha\beta}Z_1\omega  +
Y_2^{\alpha\beta}Z_2\lambda^* + Y_3^{\alpha\beta}Z_3\lambda + Z_4
Y_4^{\alpha\beta} ) \label{144} \ .
\end{eqnarray}
These are the components that enter the equations of motion.


\section{\N = 4 and \N = 3 Supersymmetric Flux Vacua}
\resetcounter


Before we start exploring different flux vacua, we should give our
index conventions: ``p--t'' denote the base directions, ``i--n''
the fiber directions, and ``a--g'' are the indices of the whole
7-d internal space.

We will now explore supersymmetric vacua  for our spinor Ans\"atze and start
with the maximal supersymmetric case, which is \N = 4 in our framework. In
this case, we decompose the gravitino mass matrix in such a way that
\begin{eqnarray}
  W = \left\{ \ba{ll} W_1 & W_3 \\
 W_2 & W_4 \ea \right \}.
\end{eqnarray}
with $W_i$ being $2 \times 2$ sub-matrices. Hence, we get
\begin{eqnarray}
  \tilde \eta &=&  (W_1^{xy} \epsilon_{xL} + W_2^{xy}
  \epsilon_{xR}^*) \otimes a^*_{yL}\theta_y^* + (
  W_{3}^{xy}\epsilon_{xL} + W_4^{xy} \epsilon_{xR}^*) \otimes a_{yR}
\theta_y  + cc
\end{eqnarray}
and the Killing spinor equation (\ref{external KQ0}) splits into
two equations, related to the opposite chirality of the external
spinors. Convenient combinations become
\begin{eqnarray}
  (W_{+}^{xy}\theta_y^*+\widetilde W_{+}^{xy}\theta_y) +
  \frac{1}{72}e^{-3U}  F
  \theta^x =0  \label{3} \\(W_{-}^{xy}\theta_y^*+\widetilde
  W_{-}^{xy}\theta_y) + (
  \partial U + \frac{im}{36}e^{-3U})\theta^x =0 \label{4}
\end{eqnarray}
with
\begin{eqnarray}
W_{\pm}^{xy}&=& (a^{x}_L)^{-1}W_1^{xy}a_{L}^{y*}\pm(a^x_{R})^{-1}
\bar
W_4^{xy}a_{R}^{y*}, \\
 \widetilde W_{\pm}^{xy}&=&
(a^x_{L})^{-1}W_3^{xy}a_R^y\pm(a^x_{R})^{-1} \bar W_2^{xy}a_L^y \
.
\end{eqnarray}
Similarly, the internal Killing spinor equation (\ref{internal
KQ0}) can be written as
\begin{eqnarray}
 \gamma_a(W_{+}^{xy}\theta_y^*+\widetilde W_{+}^{xy}\theta_y) +
\Big(\, \frac{1}{6}e^{-3U}  F_a + \partial_a \log{a_R^x \over
a_L^x} \,
\Big)
\theta_x = 0 \label{7} \ , \\
\Big[\nabla_a-\frac{ime^{-3U}\gamma_a}{48}-\frac{1}{2}\gamma_a\partial
U+\frac{1}{2} \partial_a \log{a_L^x a_R^x}
\Big]\theta^x -\frac{1}{2}
\gamma_a(W_{-}^{xy}\theta_y^*+\widetilde W_{-}^{xy}\theta_y) =0
\label{8} \ .
\end{eqnarray}
Note, there is no summation over the index $x$!

\bigskip

\noindent First, let's consider eq.\ (\ref{3}) and eq.\ (\ref{7}),
which become
\begin{eqnarray}
 (W_{+}^{xy}\theta_y^*+\widetilde W_{+}^{xy}\theta_y)+\frac{\gamma^a}{5}
 \partial_a \log {a_L^x  \over a_R^x} \,  \theta^x& =&0 \ .
\label{42}
\end{eqnarray}
The contraction with $\theta^T$ and $\theta^\dagger$ yields
\begin{eqnarray}
  W_{+}&=&0 \ , \\
 \widetilde W_{+}^{xy}+\frac{1}{5} (v^a)^{yx}\,  \partial_a
\log{ a_L^x \over a_R^x} & =&0
\end{eqnarray}
(where the forms in (\ref{SU(2) bilinears}) are used) and the
contraction with $\theta^{\dagger}\gamma^{p}$ gives
\begin{eqnarray}
\partial_p \log{ a_L^x \over a_R^x} & =&0  \label{78} \ .
\end{eqnarray}
Recall, in our notation: ``p'', ``q'' denote the base directions.
Eq.\ (\ref{42}) also implies that (\ref{7}) can be rewritten as
\begin{eqnarray}
\Big[\frac{e^{-3U}}{6}F_a - \Big(\frac{\gamma_a^b}{5} +
\frac{6\delta_a^b}{5}\Big)\, \partial_b \log{a_L^x \over a_R^x} \,
\Big] \, \theta^x &=& 0 \ . \label{122}
\end{eqnarray}
By considering eq.\ (\ref{122}) with $v^a\theta^\dagger$ and
$\theta^\dagger \gamma^a$, we obtain
\begin{eqnarray}
  A &=&0\ , \\
\widetilde W_{+}&=&\omega \haken C=(v^a)^{xy} \partial_a \log
{a_L^y \over a_R^y} =0 \label{90}
\end{eqnarray}
Now, projecting eq.\ (\ref{122}) on the base and contracting it
with $\theta^\dagger\gamma_q$, we have
\begin{eqnarray}
-i(C^{\alpha\beta})_p^{\ c}\omega_{cq} +
(C^{\alpha\beta})_{pq}&=&0
\end{eqnarray}
According to the reality conditions, no C part and hence no internal fluxes
can be turned on for the \N = 4 case.

\bigskip

\noindent Next,  the Killing spinor equations (\ref{4}) and
(\ref{8}) are explored in the same way. The contractions with
$\theta$, $\theta^\dagger$ and $\theta^\dagger  \gamma_p$, yield
\begin{eqnarray}
W_{-}&=&0 \\
 \widetilde W_{-}^{xy} & = & - v^{yx} \haken \partial U -
\frac{im}{36}e^{-3U}\delta^{xy} \label{46} \\
\partial_q U &=&0 \label{50}
\end{eqnarray}
and therefore the warp factor has to  be constant over the base. On the other
hand, combining eq.\ (\ref{4}) and eq.\ (\ref{8}) gives
\begin{eqnarray}
\Big[\, \nabla_a-\frac{ime^{-3U}\gamma_a}{144} +\frac{1}{2}
\partial_a \log ({ a_{xL} a_{xR}}) \, \Big]\theta_x &=& 0
\label{47}
\end{eqnarray}
which implies the following differential equations for the
$SU(2)$-structure
\begin{eqnarray}
  \nabla_a  \Sigma^{(k)}_{xy} & = & \theta_x^\dagger \Big
    [\gamma^{(k)},\frac{ime^{-3U}\gamma_a}{144}  \Big ]
 \theta_y  -  \frac{1}{2} \partial_a \log (a_{xL}^\star a_{xR}^\star
 a_{yL} a_{yR})\; \Sigma^{(k)}_{xy} \label{N=4 G-structure equation 1} \ , \\
\nabla_a  {\Omega}^{(k)}_{xy} & = &  \theta_x^T \Big
[\gamma^{(k)},\frac{ime^{-3U}\gamma_a}{144}  \Big ] \theta_y  -
\frac{1}{2} \partial_a \log (a_{xL}^\star a_{xR}^\star
  a_{yL} a_{yR})\; \Omega^{(k)}_{xy}  \label{N=4 G-structure equation 2} \ .
\end{eqnarray}

According to eq.(\ref{SU(2) bilinears}), constraints $\Sigma^{(0)}\equiv \I$
and $\Omega^{(0)}=\Omega^{(1)}\equiv 0$ imply that these representative
matrices satisfy\footnote{In the literature the second and third constraints
are often neglected although they impose severe constraints; see also below.}
\begin{eqnarray}
\nabla_a \Sigma^{(0)}=\nabla_a \Omega^{(0)}=\nabla_a
\Omega^{(1)}_b\equiv 0 \label{SU(2)} \ .
\end{eqnarray}
where the indices ``x'' and ``y'' are removed for simplicity. The first
condition is satisfied only if
\begin{eqnarray}
  \partial_a ( \log |a_{xL}|^2 |a_{xR}|^2 ) = 0
  \label{48}
\end{eqnarray}
and due to eq.\ (\ref{78}) and eq.\ (\ref{90}) this implies
\begin{eqnarray}
a_{xL}=e^{i\phi_{x}}\quad , \qquad a_{xR}=e^{i\phi_{x}}
\label{100}
\end{eqnarray}
up to a constant factor and $\phi_{x}=\phi_{x}(y_a)$ being real functions on
the internal coordinates. As for the other two conditions, the second one can
be automatically satisfied, however,  the third one requires
\begin{eqnarray}
m=0 \ .
\end{eqnarray}
This means that $\widetilde W_{-}$ is only associated with the
fiber dependence of the warped factor according to eq.\
(\ref{46}).

 From the $SU(2)$-structure differential equations we can also infer the
geometry of the internal space. First, consider the differential equation of
the vectors, which becomes after taking into account (\ref{48})
\begin{eqnarray}
\nabla_a  v^\alpha_b(\sigma_\alpha)_{xy}
 & =& -  \frac{i}{2} \partial_a\Big( \log\frac{a_{1L}a_{1R}}{a_{2L}a_{2R}}\Big)
(v^2_b \sigma_1 + v^1_b \sigma_2)_{xy} \label{114}
\end{eqnarray}
and the antisymmetrization gives
\begin{eqnarray}
(d v^1+i d v^2 )_{ab}  &=&  -(v^1 - i v^2)_{[a}\partial_{b]}\log
\frac{a_{1L}a_{1R}}{a_{2L}a_{2R}} \ , \\
(d v^3)_{ab} &=& 0\label{N=4 v-equation (anti)} \ .
\end{eqnarray}
These global vector fields can be Killing, $i.e.$,
$\nabla_{\{a}v^\alpha_{b\}}\equiv 0$, only if $\phi_1=\phi_2$. The
differential equation of the almost complex structure $\omega$ (of
the base) reads
\begin{eqnarray}
 \nabla^a (\Sigma^{(2)}_{bc})_{xy}
 &=&
 -  \frac{1}{2}
\partial_a \log (a_{xL}^\star a_{xR}^\star
  a_{yL} a_{yR}) (i\omega_{bc}+(v\wedge v)_{bc})_{xy}
 \label{49}
\end{eqnarray}
and with (\ref{48})  we find
\begin{eqnarray}
d \omega &=& 0 \ .
\end{eqnarray}
The differential equation for $\lambda^*$ becomes
\begin{eqnarray}
\nabla^a (\Omega^{(2)}_{ab})_{xy}  &=& \frac{1}{2}
\partial_a \log( a_{xL} a_{xR} a_{yL} a_{yR}) \;
  \lambda^*_{bc} (\sigma_2)_{xy}
\end{eqnarray}
which leads
\begin{eqnarray}
(d\lambda^*)_{abc} &=& -\frac{3}{2} \partial_{[a} \log (a_{1L}
a_{1R} a_{2L} a_{2R}) \lambda^*_{bc]} \label{59}
\end{eqnarray}
Therefore the 4-d base space is conformal to a K\"ahler space and
becomes hyper K\"ahler if the two phase a $a_1$ and $a_2$ cancel,
ie.\ $\phi_1+\phi_2=0$.

In summary, we only find trivial flux vacua for the ${\cal N}=4$
case, with the external space-time being flat and the base
manifold of the internal space being K\"ahler. Actually, this
vacuum admits $SU(2)$ holonomy.

\bigskip

\noindent Finally, let us comment on the \N = 3 case.  With the
projector (\ref{03}) the $4\times 4$ \N = 4 gravitino mass matrix
is reduced to a $3\times 3$ matrix
\begin{eqnarray}
  W = \left\{ \ba{ll} W_1 & (W_3)^{x1} \\
 (W_2)^{1y} & (W_4)^{11} \ea \right \} \ ,
\end{eqnarray}
which fixes the spinor $\tilde \eta$ as
\begin{eqnarray}
\tilde \eta &=&   (W_1^{xy} \epsilon_{xL} + W_2^{1y}
\epsilon_{1R}^*) \otimes \chi_{yL}^* +  ( W_{3}^{x1}\epsilon_{xL}
+ W_4^{11} \epsilon_{1R}^*) \otimes \chi_{1R}  + cc
\end{eqnarray}
the \N = 4 external Killing spinor equations  are truncated to
\begin{eqnarray}
  (W_1^{xy} \chi^*_{yL}+W_3^{x1} \chi_{1R}) + (\frac{1}{2}
\partial U + \frac{im}{72}e^{-3U} + \frac{1}{144}e^{-3U}  F ) \ ,
\chi^x_L =0 \label{11}\\
  ( \bar W_2^{1y} \chi_{yL} +\bar W_4^{11} \chi_{1R}^* ) +
(-\frac{1}{2}
\partial U - \frac{im}{72}e^{-3U} + \frac{1}{144}e^{-3U}  F ) \ .
\chi_R^1  =0 \label{12}
\end{eqnarray}
Similarly, the \N = 4 internal Killing spinor equations  are
reduced to
\begin{eqnarray}
\nabla_a \chi^x_L &=&  \gamma_a (W_1^{xy} \chi^*_{yL}+W_3^{x1}
\chi_{1R}) \nonumber
\\&&+\Big(\frac{e^{-3U}}{12}
F_a+\frac{1}{2}\gamma_a \partial U
+\frac{ime^{-3U}\gamma_a}{48}\Big)
\chi^x_L \label{224} \ , \\
\nabla_a \chi_R^1&=&- \gamma_a ( \bar W_2^{1y} \chi_{yL} +\bar
W_4^{11} a_{1R}^*\theta^*_1 )\nonumber
\\&&+\Big(-\frac{e^{-3U}}{12}F_a+\frac{1}{2}\gamma_a \partial U+
\frac{ime^{-3U}\gamma_a}{48}\Big)\chi_R^1 \label{225}
\end{eqnarray}
where we used the notation: $\chi_{1L}=a_{1L} \theta_1$ , $
\chi_{2L} = a_{2L}\theta_2+ a_{2R}^*\theta_2^* $ and
$\chi_{1R}=a_{1R} \theta_{1R}$.


\section{\N = 2 Supersymmetric Flux Vacua}
\resetcounter

We turn now to the more interesting case of supersymmetric flux vacua with \N
= 2 supersymmetry, which can have $SU(2)$ or $SU(3)$ structures.  We will
treat each case separately.

\subsection{$SU(2)$ Structures}


Here, we have to take the spinor Ansatz
(\ref{N=2(I)}) and gravitino masses matrix is projected to
\begin{eqnarray}
W&=&W_1
\end{eqnarray}
which is the  $2\times 2$ sub-matrix of the \N = 4 one defined
before. With
\begin{eqnarray}
\tilde \eta &=&   W^{xy} \epsilon_{xL}  \otimes \chi_{y}^* +
  \bar W^{xy} \epsilon_{xL}^*  \otimes
\chi_y
\end{eqnarray}
the  Killing spinor equations are truncated to
\begin{eqnarray}
0&=&  W^{xy} \chi^*_{y} + \Big( \, \frac{1}{2}
\partial U + \frac{im}{72}e^{-3U} + \frac{1}{144}e^{-3U}  F \, \Big) \,
\chi^x  \label{07} \ , \\
\nabla_a \chi^x &=&  \gamma_a W^{xy} \chi^*_y
+\Big(\frac{e^{-3U}}{12} F_a+\frac{1}{2}\gamma_a \partial U
+\frac{ime^{-3U}\gamma_a}{48}\Big)\chi^x \label{08}
\end{eqnarray}
where we used the notation
\be211
\chi_{1}=a_{1L} \theta_1 + a_{1R}^*\theta_1^* \quad , \qquad \chi_{2} =
a_{2L}\theta_2+ a_{2R}^*\theta_2^* \label{06} \ .
\ee
These are the most general \N = 2 Killing spinor equations, and all flux
solutions with $SU(2)$ structures should satisfy these two equations. We will
not discuss this most general case. We shall  instead explore the three
special cases that we mentioned after eq.\ (\ref{N=2(I)}).


\subsubsection{Case (a)}


In this case $\chi_x$ are truncated to
\begin{eqnarray}
\chi_{1}=a_{1L} \theta_1 \quad , \qquad \chi_{2} = a_{2L}\theta_2
\end{eqnarray}
and the Killing spinor equations become
\begin{eqnarray}
0&=&  W^{xy}a_{y}^* \theta^*_y + (\frac{1}{2}
\partial U + \frac{im}{72}e^{-3U} + \frac{1}{144}e^{-3U}  F ) \,
a^x \theta^x  \label{131} \ , \\
\nabla_a \theta^x  &=& \gamma_a (a^x)^{-1} W^{xy}a_y^*\theta_y^*
+ \Big( \frac{e^{-3U}}{12}
F_a+\frac{1}{2}\gamma_a\partial U- \frac{\partial_a a^x}{a^x}
+\frac{ime^{-3U}\gamma_a}{48}\, \Big)\theta^x \label{132}
\end{eqnarray}
where suppressed  the subscripts ``L'' and ``R''.  Contracting eq.\
(\ref{131}) with $\theta^T$ and $\theta^\dagger$ gives
\begin{eqnarray}
a_x^*W^{xy}&=&\frac{e^{-3U}a_{x}}{24}((C \haken \lambda^*)
 \sigma^2)_{xy}
\label{contraction 1}  \ , \\
\partial U\haken v&=&
-\frac{im}{36}e^{-3U}-\frac{e^{-3U}}{72}(6i C \haken \omega
-A\haken vol_4) \label{contraction 2}
\end{eqnarray}
which implies
\begin{eqnarray}
\frac{im}{36}e^{-3U}=A\haken vol_4 &=&0 \label{contraction 2a}\ , \\
\partial U\haken v &=&
-\frac{ie^{-3U}}{12} C \haken \omega\label{contraction 2b} \ .
\end{eqnarray}
{From} the contraction with $\theta^T\gamma_p$ or
$\theta^{\dagger}\gamma_p$, we get
\begin{eqnarray}
\partial_p U =0 \ .
\end{eqnarray}
Recall, we consider only regular fluxes, ie.\ we do not take into
account any fluxes which are 3- or 1-forms on the base.

\bigskip

\noindent
The internal Killing spinor equation (\ref{132}), yields again
differential equations for the $G$-structures, which we have given
in the Appendix. The self-consistence of $SU(2)$-structures requires
\begin{eqnarray}
\nabla_a \Sigma^{(0)}=\nabla_a \Omega^{(0)}=\nabla_a
\Omega^{(1)}_b=0 \ .
\end{eqnarray}
The first condition leads $\partial_a U = \partial_a \log|a_x|^2$
or
\begin{eqnarray}
  a_x=e^{U/2+i \, \phi_x} \label{135} \ .
\end{eqnarray}
Here $\phi_x$ is a real function on the internal coordinates.  With
(\ref{contraction 1}), we find from the second condition
\begin{eqnarray}
C_{\alpha\beta}\haken \lambda^*=0 \quad , \qquad W = 0 \ .
\end{eqnarray}
The last condition is automatically satisfied.
The differential equations for the vector becomes
\begin{eqnarray}
\nabla_a v^\gamma_b &=& \frac{-e^{-3U}}{6} F_a^{\ cde}
(v_\alpha\wedge v_\beta\wedge \omega)_{bcde}
\epsilon^{\alpha\beta\gamma} + \frac{1}{2}\partial^c
U\delta_{b[a}v_{c]}^\gamma + \Delta^\gamma_{ab} \label{333}
\end{eqnarray}
with $ \Delta^\gamma_{ab} (\sigma_\gamma) = i \partial_a \log( {a_{1}
\over a_2}) (v^2_b \sigma_1 + v^1_b \sigma_2)$.  With the notation
(\ref{141}), the antisymmetrization gives
\begin{eqnarray}
(d v^\gamma)_{ab}  = 32 \epsilon_{\alpha\beta}^{\ \ \gamma}
Y_1^{\beta\lambda}(v_\lambda)_{[a}v^\alpha_{b]} - v_{[a}^\gamma
\partial_{b]} U  + 2\Delta^\gamma_{[ab]}\label{320} \ .
\end{eqnarray}
{From} (\ref{G-structure equation 1}) with $W=0$, we find
\begin{eqnarray}
  \nabla_a \omega_{bc}&=&-2(\delta_{a[b}\omega_{c]d}-
  \delta_{d[b}\omega_{c]a})\partial^d U \label{327}
\end{eqnarray}
and hence
\begin{eqnarray}
(d e^{-6U} \omega)_{abc} & = &  0 \label{328} \ .
\end{eqnarray}
With  (\ref{135}) we get for $\lambda$
\begin{eqnarray}
\nabla_a \lambda_{bc} &=& \partial_a \log {a_1 \over a_2^\star} \;
 \lambda_{bc}-2(\delta_{a[b}\lambda_{c]d}-\delta_{d[b}(\lambda_{c]a})
 \partial^d
 U\label{331}
\end{eqnarray}
and anti-symmetrizing this equation yields
\begin{eqnarray}
(d \lambda)_{abc} &=& 3\partial_{[a} (\log\frac{a_1}{a_2^*})
\lambda_{bc]} +
 6\lambda_{[ab}\partial_{c]} U  \label{332} \ .
\end{eqnarray}
So, the 4-d base space is conformal to a K\"ahler space, which
becomes hyper-K\"ahler (ie.\ all three 2-forms are closed) if
$\phi_1 + \phi_2 = 0$.

In summary, we find flux vacua, satisfying Killing spinor
equations,  where $[C]: 1+3$ components can be turned on, with the
external space-time being flat and the base manifold of the
internal space being K\"ahler. These vacua are allowed to be
warped along fiber direction which is mediated by the $1$
component.

\bigskip
\noindent
Finally, we consider the Bianchi identities and equations of motion.
The only non-trivial flux components (and their Hodge dual) are
\begin{eqnarray}
F& =& e^{3U}(v_{\alpha}\wedge v_{\beta}) \wedge
(Y^{\alpha\beta}_1\omega +Y_4^{\alpha\beta}) \\
{*} F&=& e^{3U}\sqrt{G_7}\, \epsilon_{\alpha\beta\gamma} v^{\gamma}
\wedge (Y_1^{\alpha\beta}Z_1\omega  + Z_4Y_4^{\alpha\beta} )
\end{eqnarray}
where we use the notation as introduced at eq.\ (\ref{141}).
Hence, the Bianchi identity and the equation of motion become
\begin{eqnarray}
0= e^{-3U}(d F)_7
&=& 2(dv_\alpha \wedge v_\beta) \wedge (Y^{\alpha\beta}_1\omega +
Y_4^{\alpha\beta}) \nonumber \\
&& + (v_\alpha \wedge v_\beta) \wedge (d(Y^{\alpha\beta}_1\omega)+
 dY_4^{\alpha\beta}) \nonumber \\
&&+ 3  dU \wedge (v_{\alpha}\wedge v_{\beta}) \wedge
(Y^{\alpha\beta}_1\omega + Y_4^{\alpha\beta}) \label{155} \ , \\
0= \frac{e^{-7U}d [ e^{4U} ({*} F)_7]}{\sqrt{G_7}}
&=&\epsilon_{\alpha\beta\gamma} dv^{\gamma} \wedge
(Y_1^{\alpha\beta}Z_1\omega  +
Z_4Y_4^{\alpha\beta} ) \nonumber \\
&& -\epsilon_{\alpha\beta\gamma} v^{\gamma} \wedge
[d(Y_1^{\alpha\beta}Z_1\omega)   +
d(Z_4Y_4^{\alpha\beta}) ] \nonumber \\
&&+ 14\,  \epsilon_{\alpha\beta\gamma} dU\wedge v^{\gamma} \wedge
(Y_1^{\alpha\beta}Z_1\omega   + Z_4Y_4^{\alpha\beta} )
\label{156} \ .
\end{eqnarray}
These two equations are not easy to solve, but let us consider some
simplified cases. If the {\bf 1} component of $C$ is zero,
eq.\ (\ref{contraction 2b}) gives $dU=0$ and if we set moreover
$\partial_{a}(\phi_1-\phi_2)=0$,  then one finds
\begin{eqnarray}
dv=d\omega=0 \quad &, & \qquad
(d \lambda)_{abc} = 3\partial_{[a} (\log\frac{a_1}{a_2^*})
\lambda_{bc]}
\end{eqnarray}
and (\ref{155}), (\ref{156}) are simplified to
\begin{eqnarray}
0&=&  v_\alpha \wedge v_\beta \wedge  dY_4^{\alpha\beta} \label{231} \ ,\\
0&=& -\epsilon_{\alpha\beta\gamma} v^{\gamma} \wedge (dZ_4\wedge
Y_4^{\alpha\beta}+Z_4 dY_4^{\alpha\beta})
= {*} v_\alpha \wedge v_\beta \wedge d^\dagger Y_4^{\alpha\beta}
\label{232} \ .
\end{eqnarray}
These equations are solved if the 2-forms $Y_4^{\alpha\beta}$ are
harmonic and $Z_4$ is constant.  Thus, the Bianchi identity and the
equation of motion can be solved if the ${\bf 3}$ components of the $C$
flux are harmonic and the other fluxes are trivial.


\subsubsection{Case (b)}


Here, we consider
\begin{eqnarray}
\chi_{1}= a_{1L} \theta_1 \quad , \qquad \chi_{2}& =& a_{2R}^*\theta_2^*
\end{eqnarray}
and the Killing spinor equations  (\ref{07}) and  (\ref{08})
become
\begin{eqnarray}
  (W_{11}a_{1}^* \theta_1^*+W_{12}a_2 \theta_2)+
(\frac{1}{2}
\partial U + \frac{im}{72}e^{-3U} + \frac{1}{144}e^{-3U}  F ) \,
a_1 \theta_1 =0 \ , \label{201}\\
  (\bar W_{21}a_{1}\theta_1+\bar W_{22}a_2^* \theta_2^*) +
(-\frac{1}{2}
\partial U - \frac{im}{72}e^{-3U} + \frac{1}{144}e^{-3U}  F ) \,
a_2 \theta_2 =0 \label{202}
\end{eqnarray}
and
\begin{eqnarray}
\nabla_a \theta_1 &=&  \gamma_a (\frac{a_1^*}{a_1} W_{11}
\theta_1^*+\frac{a_2}{a_1}W_{12}
\theta_2)+\nonumber \\
&&(\frac{e^{-3U}}{12} F_a+ \frac{1}{2}\gamma_a\partial U-
\frac{\partial_a a_1}{a_1} +\frac{ime^{-3U}\gamma_a}{48})\theta_1
\ ,\label{203}\\
\nabla_a \theta_2 &=& - \gamma_a ( \frac{a_1}{a_2}\bar
W_{21}\theta_1+\frac{a_2^*}{a_2}\bar W_{22} \theta_2^*)+
\nonumber \\
&& (-\frac{e^{-3U}}{12}F_a+\frac{1}{2}\gamma_a\partial
U-\frac{\partial_a a_2}{a_2}+\frac{ime^{-3U}\gamma_a}{48})\theta_2
\label{204} \ .
\end{eqnarray}
Again for simplification we have changed the indices in the Killing
spinor Ansatz (\ref{N=2(I)b}) so that
\begin{eqnarray}
\eta &=& a_{1}\epsilon_L\otimes \theta_1 + a_{2}\epsilon_R \otimes
\theta_2 +cc \ .
\end{eqnarray}

\bigskip
\noindent Severe constraints come again from the consistency requirements
\begin{eqnarray}
\nabla_a \Sigma^{(0)}=\nabla_a \Omega^{(0)}=\nabla_a
\Omega^{(1)}_b=0 \label{SU(2) consistence} \ .
\end{eqnarray}
For the second and third condition,
(\ref{G-d equation7}) and (\ref{G-d equation8}) require
\be721
W_{11}=  W_{22} = m = C_a^c \lambda^*_{cb} =0
\ee
and thus, the Freund-Rubin and the C part fluxes have to vanish.
Contracting eq.\ (\ref{201}) and eq.\ (\ref{202}) with
$\theta_x^T\gamma_p$ or $\theta_x^{\dagger}\gamma_p$, we obtain
\begin{eqnarray}
\partial_p U =0
\end{eqnarray}
and from the $\theta_x^\dagger$ contraction we find
\bea322
\frac{a_2}{a_1}\, W_{12}&=&-(\frac{1}{2}\partial
U\haken v)_{21} = -
\frac{a_1^\star}{a_2^\star} W_{21} \ , \\
A\haken vol_4 &=& 72 e^{3U}(\partial U \haken v)_{11}, \label{82}
\eea
Using these results, we find for $\nabla_a \Sigma^{(0)}=0$ as
non-trivial conditions
\begin{eqnarray}
\nabla_a  \Sigma^{(0)}_{11} & = & - \partial_a \log(e^{-U} |a_1|^2) +
 \Big(\frac{a_2^*}{a_1^*}\bar{W}_{12}(v_a)_{21}+\frac{a_2}{a_1}
W_{12}(v_a)_{12}\Big)=0  \label{norm 1} \\
\nabla_a  \Sigma^{(0)}_{22} & = &
 - \partial_a \log(e^{-U} |a_2|^2)-
 \Big(\frac{a_1^*}{a_2^*}W_{21}(v_a)_{12}+\frac{a_1}{a_2}\bar
W_{21}(v_a)_{21}\Big)=0  \label{norm 2}
\end{eqnarray}
which leads to while using (\ref{322})
\begin{eqnarray}
&& \partial_a\Big(\log \frac{|a_1|}{|a_2|}\Big) = 0  \\
&& \partial_a[\log (e^{-2U}|a_1|^2|a_2|^2)] = \frac{a_2}{a_1}
W_{12}(v_a)_{12}+ c.c. \label{881}
\end{eqnarray}
So far, all of our discussions are based on a general background.
For a Minkowski vacuum we have to ensure that all components of
$W$ are zero and hence
\begin{eqnarray}
a_x & \sim &
e^{U/2+i \phi_x}\\
\partial_1 U &=& \partial_2 U =0 \label{234} \\
A\haken vol_4&= &72 e^{3U}\partial_3 U \haken v^3 \label{83}
\end{eqnarray}
For the derivative of the vectors $v^\alpha$ we  now find
\begin{eqnarray}
(d v^1)_{ab} & = & -2iv_{[a}^2\partial_{b]}\log\frac{a_1}{a_2}+4
v_{[a}^1
\partial_{b]}U \nonumber \\
(d v^2)_{ab} & = &-2iv_{[a}^1\partial_{b]}\log\frac{a_1}{a_2}+4
v_{[a}^2
\partial_{b]}U
\nonumber \\ (d v^3)_{ab} & = & 4 v_{[a}^3
\partial_{b]}U\label{341}
\end{eqnarray}
and in addition
\begin{eqnarray}
d (e^{-6U} \omega) & = &  0  \label{342}
\end{eqnarray}
and hence
\begin{eqnarray}
(d \lambda)_{abc} &=& 3\partial_{[a} (\log\frac{a_1}{a_2^*})
\lambda_{bc]} + 6\lambda_{[ab}\partial_{c]} U
\end{eqnarray}
Therefore the 4-d base space is conformal to a K\"ahler space and
becomes hyper K\"ahler if the two phase a $a_1$ and $a_2$ cancel,
ie.\ $\phi_1+\phi_2=0$.

In summary, we find flux vacua,  satisfying  Killing spinor equations, where
only A part fluxes can be turned on, with the external space-time being flat
and the base manifold of the internal space being K\"ahler. These vacua are
allowed to be warped along a fiber direction which is mediated by the A part
flux as $SU(2)$ singlet.

\bigskip
\noindent Again, at the end we want to discuss  Bianchi identities and
equations of motion. Since in this case  only the $A$ part of the flux can be
turned on, the flux decomposition \ (\ref{141}) and its Hodge dual \
(\ref{144}) are reduced to
\begin{eqnarray}
F& =& e^{3U}X \, (\omega \wedge \omega) \label{346} \\
{*} F&=& e^{3U}\sqrt{G_7} X Z \, \epsilon_{\alpha\beta\gamma} v^{\alpha}
\wedge v^{\beta} \wedge v^{\gamma} \label{347}
\end{eqnarray}
and with (\ref{342}), the Bianchi identity  and the equations of motion lead
to
\begin{eqnarray}
0&=&(dX + 7 X dU) \wedge \omega \wedge \omega  \label{344}\\
0&=& \epsilon_{\alpha\beta\gamma}(14 XZ dU + d(XZ))\wedge
v^{\alpha} \wedge v^{\beta} \wedge v^{\gamma} \nonumber \\&& + 3
XZ \epsilon_{\alpha\beta\gamma}dv^{\alpha} \wedge v^{\beta} \wedge
v^{\gamma} \label{345}
\end{eqnarray}
For the Minkowski case with $\phi_1=\phi_2$ or
$(dv^\gamma)_{ab}=4v^\gamma_{[a}\partial_{b]}U$, eq.\ (\ref{345}) gives
\begin{eqnarray}
XZ\propto e^{-20 \, U}
\end{eqnarray}
In this case the Bianchi identity is rather constrained. It implies $X\propto
e^{-7U}$ and hence contradicts with Killing spinor equations or eq.\
(\ref{82}). However, we would point out again that this can be remedied by
adding sources; namely, non-vanishing $dF$ could be turned on in the presence
of M-brane sources.


\subsubsection{Case (c)}


Now the 11-d Majorana spinor is written as
\begin{eqnarray}
\eta &=& a_{1}\epsilon_{1}\otimes \theta_1 + a_{2}\epsilon_{2}
\otimes (\theta_2+\theta_2^*) +cc\ .
\end{eqnarray}
We repeated the calculations in the same way as for Case (a) and (b) and found
many constraints, however not explicit results for interesting vacuum
solutions. Here we shall  not present  explicit (lengthy) calculations and
shall instead turn to the case with $SU(3)$ structures.


\subsection{$SU(3)$ Structures} \label{N=2 Flux(I)}


In this case we have only one internal spinor $\theta$ and two
external spinors. Hence, the gravitino mass matrix is projected to
\begin{eqnarray}
  W =\left\{ \ba{ll} W^{11} & W^{12} \\
 W^{21} & W^{22} \ea \right \}= \left\{ \ba{ll} (W_1)^{11} & (W_3)^{11} \\
 (W_2)^{11} & (W_4)^{11} \ea \right \} \ ,
\end{eqnarray}
which fixes the spinor $\tilde \eta$ as follows
\begin{eqnarray}
  \tilde \eta &=&   (W^{11}\epsilon_L+
  W^{21}\epsilon_R^*)\otimes a_L^*\theta^*
  + (W^{12}\epsilon_L+ W^{22}\epsilon_R^*)\otimes
  a_R\theta + cc
\end{eqnarray}
From the external as well as from internal Killing spinor
equations we get two equations, which are related to the opposite
chirality of the external spinor. With the appropriate
combinations these two sets of equations become
\begin{eqnarray}
  (W_{A+}\theta^*+W_{B+}\theta) + \frac{1}{72}e^{-3U} F
  \theta =0  \label{105} \ , \\
  (W_{A-}\theta^*+W_{B-}\theta) + (
  \partial U + \frac{im}{36}e^{-3U})\theta =0 \label{106}
\end{eqnarray}
and
\begin{eqnarray}
  \gamma_a(W_{A+}\theta^*+W_{B+}\theta) +
  \Big(\frac{1}{6}e^{-3U}  F_a
  - \partial_a \log {a_L \over a_R} \, \Big) \theta =   0 \label{107}\ , \\
  \Big[\nabla_a-\frac{ime^{-3U}\gamma_a}{48}-\frac{1}{2}\gamma_a \partial
    U+\frac{1}{2} \partial_a \log (a_L a_R)\, \Big]\theta
  -\frac{1}{2} \gamma_a(W_{A-}\theta^*+W_{B-}\theta) =0
  \label{108}
\end{eqnarray}
where we defined
\begin{eqnarray}
 W_{A\pm}&=& \frac{a_L^*}{a_L} W^{11} \pm  \frac{a_R^*}{a_R}\bar W^{22}\ , \\
  W_{B\pm}&=& \frac{a_R}{a_L}W^{12} \pm \frac{a_L}{a_R}\bar W^{21} \ .
\end{eqnarray}
The constraint equation (\ref{107}) can be used to eliminate the
flux part in (\ref{105}) and the contraction with $\theta$ and
$\theta^\dagger$ yields
\begin{equation}
W_{A+}=0 \quad , \qquad W_{B+}=\frac{v^a}{5}\Big(\frac{\partial_a
a_L}{a_L}-\frac{\partial_a a_R}{a_R}\Big) \label{93} \ee
and from the contraction with $\theta^\dagger \gamma^p$ we find
\begin{eqnarray}
  \partial_p \log {a_L \over a_R} = 0 \ .
\end{eqnarray}
Therefore, by contracting (\ref{107}) with $\theta$ and
$\theta^\dagger$, the $p$-component gives
\begin{eqnarray}
(G \haken \Psi)_p =0 \quad &, & \qquad  (H\haken J)_p=0 \ .
\end{eqnarray}
With this result one finds moreover
\begin{eqnarray}
  d \log {a_L \over a_R} = 0 \label{96}
\end{eqnarray}
and hence $W_{B+} = 0$ and  in addition
\begin{eqnarray}
F_a \theta=0 \label{24} \ .
\end{eqnarray}
It is straightforward to show (by repeating all possible
contractions) that this equations imply that all internal fluxes
have to vanish.

\bigskip
\noindent It remains to  explore eq.\ (\ref{106}) and the differential
equation (\ref{108}). The contraction with $\theta^T$ and $\theta^\dagger$
gives
\begin{equation}
W_{A-}=0 \ , \qquad W_{B-}=- v \haken \partial U -
\frac{im}{36}e^{-3U} \label{68} \ee
and the contraction with $\theta^\dagger \gamma^p$ yields
\begin{eqnarray}
  \partial_q U=0 \ .
\end{eqnarray}
On the other hand, combining eq.\ (\ref{106}) and eq.\
(\ref{108}), we have
\begin{eqnarray}
\Big[\nabla_a-\frac{ime^{-3U}\gamma_a}{144} +\frac{1}{2}
\partial_a \log (a_L a_R) \, \Big]\theta &=& 0 \label{44}
\end{eqnarray}
and hence the differential equations for the $G$ structures become
\begin{eqnarray}
\nabla_a  \Sigma^{(k)} & = & \theta^\dagger \Big
[\gamma^{(k)},\frac{ime^{-3U}\gamma_a}{144}  \Big ]
 \theta
 -   \partial_a \log(|a_L||a_R|) \,
\Sigma^{(k)}
  \label{N=2(I) G-structure equation 1} \ , \\
\nabla_a  {\Omega}^{(k)} & = &  \theta^T \Big
[\gamma^{(k)},\frac{ime^{-3U}\gamma_a}{144}  \Big ] \theta
  - \partial_a \log (a_L a_R)
\, \Omega^{(k)}   \label{N=2(I) G-structure equation 2} \ .
\end{eqnarray}
Because
\begin{eqnarray}
\nabla_a \Sigma^{(0)}=\nabla_a \Omega^{(0)}=\nabla_a
\Omega^{(1)}_b=\nabla_a\Omega^{(2)}_{bc}\equiv 0 \label{SU(3)
consistence1}
\end{eqnarray}
we infer from the first constraint
\begin{eqnarray}
 \partial_a \log(|a_L||a_R|) &=& 0
\label{partial warped factor 0} \ .
\end{eqnarray}
Together with (\ref{96}) this implies $a_L=a_R=e^{i\psi}$, up to a constant
factors. The other three equations do not impose additional constraints.
Next, for the vector field we find
\begin{eqnarray}
\nabla_a  v_b & = & \frac{m e^{-3U}J_{ab}}{72}
\end{eqnarray}
and therefore, $v$ has to be a Killing vector. In addition, we get
\begin{eqnarray}
\nabla_a J_{bc}
 &=&\frac{-m}{36}e^{-3U}\delta_{a[b} v_{c]} \label{61}
\end{eqnarray}
and hence
\begin{eqnarray}
d J=0\label{63} \ .
\end{eqnarray}
Finally, the 3-form obeys
\begin{eqnarray}
  \nabla_a \Psi_{bcd}= -i\nabla_a \Omega^{(3)}_{bcd}
  & = &  \frac{-m}{72}e^{-3U} (v \wedge  \Psi)_{abcd}  -2i
  \frac{\partial_a a_L}{a_L} \Psi_{bcd}
\end{eqnarray}
and thus
\begin{eqnarray}
  d \Psi & = &  \frac{-m}{18} e^{-3U}v \wedge
  \Psi + 8i  \Psi \wedge d \log a_L \label{64} \ .
\end{eqnarray}
So, the torsion of $X_6$ is
\be612 \tau \in {\cal W}_5 \ee
This implies that $X_6$ is always K\"ahler with the phase $\psi$
of $a_{L,R}$ as the K\"ahler connection and since the vector $v$
is Killing, the 7-d internal space is Einstein-Sasaki.  Recall,
the Freund-Rubin parameter is non-zero, but no internal fluxes are
allowed. This is a well-known vacuum of M-theory compactification
and hence there is no need to discuss the equations of motion.

In summary, we find non-warped flux vacua where only the Freund-Rubin
parameter is allowed;  the external space-time is AdS,  and the internal space
is  Einstein-Sasaki with the base manifold being K\"ahler.


\section{\N = 1 Supersymmetric Flux Vacua}
\resetcounter

\subsection{SU(3) Structures}


For the \N = 1, we have only one external spinor and hence
gravitino mass matrix is reduced to a single element, the
superpotential
\begin{eqnarray}
W=W_{11}
\end{eqnarray}
and therefore
\begin{eqnarray}
  \tilde \eta &=&   W \epsilon  \otimes \chi^* +
   \bar W \epsilon^*  \otimes \chi \ .
\end{eqnarray}
The external and internal Killing spinor equations
are now
\begin{eqnarray}
  0&=& W \chi^* + (\frac{1}{2}
  \partial U + \frac{im}{72}e^{-3U} + \frac{1}{144}e^{-3U}  F ) \,
  \chi  \label{380} \ , \\
  \nabla_a \chi &=&  \gamma_a W \chi^*
  +\Big(\frac{e^{-3U}}{12} F_a+\frac{1}{2}\gamma_a \partial U
  +\frac{ime^{-3U}\gamma_a}{48}\Big)\chi \label{381}
\end{eqnarray}
where we use the notation
\begin{eqnarray}
  \chi&=&a_{L} \theta + a_{R}^*\theta^*  \label{382} \ .
\end{eqnarray}
This spinor is normalized as $\chi^T \chi = 2 a_L a_R$ and
$\chi^\dagger \chi = a_+^2$ with
\begin{eqnarray}
a_+^2=|a_L|^2+|a_R|^2 \quad , \qquad   a_-^2=|a_L|^2-|a_R|^2 \label{334} \ .
\end{eqnarray}
Using (\ref{381}) we find therefore
\begin{eqnarray}
  \nabla_a(\chi^\dagger \chi) &\equiv& \partial_a  a_+^2 =a_+^2
  \partial_a U \label{522} \ , \\
  \nabla_a(\chi^T \chi) &\equiv&  2\partial_a  (a_La^*_R) = -2
  a_-^2Wv_a+\frac{1}{6} e^{-3U} \chi^T F_a \chi + 2a_La_R^*
  \partial_a U \label{524a}
\end{eqnarray}
which is solved by
\begin{eqnarray}
a_+^2&=&e^U \label{523}  \ , \\
\partial_a(e^{2U}a_La_R^*)&=& 2  a_-^2Wv_a \label{524}
\end{eqnarray}
and we used
\be299
  0 = -  a_-^2W v_a+\frac{1}{36} e^{-3U} \chi^T F_a \chi + a_La_R^*
\partial_a U
\ee
which is obtained by contracting (\ref{380}) with $\chi^T \gamma_a$. Let us
note, that the constraints $\nabla_a(\chi^T \gamma_a \chi) = \nabla_a(\chi^T
\gamma_{ab} \chi) = 0$ are automatically satisfied. Before we continue with
the general discussion, let us first address  special cases.


\subsubsection{Case (I)}


We assume here that
\begin{eqnarray}
\chi = a \theta \qquad {\rm or} \qquad
a_R=0 \quad , \quad a_L \equiv a\label{359}
\end{eqnarray}
which has been extensively discussed before
\cite{0303127,0311119,0311146}. We infer immediately from
(\ref{524}) and (\ref{299}) that
\be820
W\  =\  H\haken \Psi\  =\  (G \haken \psi)_p\  = \ 0
\ee
and hence there is no AdS vacuum possible in this case.  The
contraction of (\ref{380}) with $\theta^\dagger$ yields moreover
\begin{eqnarray}
m & = & 0  \ , \\
\partial_a U&=& \frac{e^{-3U}}{144} \Big( \, G \haken (J\wedge J) \, v_a
+ 4 \, (H\haken(J\wedge J))_a \, \Big) \label{con 4} \ .
\end{eqnarray}
The internal Killing spinor equation therefore  simplifies:
\begin{eqnarray}
\nabla_a \theta =  \Big( \, \frac{e^{-3U}}{12} F_a+
\frac{1}{2}\gamma_a\partial U-
\frac{\partial_a a}{a} \Big) \, \theta \ .
\label{302}
\end{eqnarray}
This yields for the $G$-structure differential equations
\begin{eqnarray}
(dv)_{ab} &=&4 v_{[a} \partial_{b]} U \label{340}
\ , \\
(dJ)_{abc}  &=&  -3e^{-3U} (J \haken G)_{[ab}v_{c]}+6e^{-3U}
H_{d[ab} J_{c]}^{\ d} + 6 \partial_{[a}U J_{bc]} \label{353}
\ , \\
(d\Psi)_{abcd}&=&-12e^{-3U} H_{f[ab} \Psi_{cd]}^{\ \ f}
 -12 \Psi_{[abc}\partial_{d]}U +
4\partial_{[a} \log \frac{a^*}{a}\Psi_{bcd]} \label{358} \ .
\end{eqnarray}
In eq.\ (\ref{358}) $H_{f[ab} \Psi_{cd]}^{\ \ f} \propto {\cal
V}_{[a}  \Psi_{bcd]}$ since $H$ has no components of $(3,0)+(0,3)$
type and the vector ${\cal V}_a$ contributes to the torsion class
${\cal W}_5$.  When projecting these expressions on $X_6$, it is
not hard to find the non-trivial torsion classes (cp.\
(\ref{192}) and (\ref{193}))
\begin{eqnarray}
  \tau\  \in\  {\cal W}_3 \oplus {\cal W}_4 \oplus {\cal W}_5
\end{eqnarray}
and therefore the 6-d base is a complex manifold and since ${\cal
W}_4$ is exact, it is in fact a so-called conformally balanced
manifold, see also \cite{0211118}.  It becomes K\"ahler only if
${\cal W}_3=0$, $i.e.$, $H^{6+\bar 6}$ is turned off. Note, these
results are identical with the those obtained in
\cite{0303127,0311146} up to numerical factors.

In summary, we find flux vacua, satisfying the Killing spinor
equations, where only $[G]: {\bf 3+ \bar 3}$ and $[G]: {\bf 1+ \bar
1}$ components are not allowed, with the external space-time being
flat and the base manifold of the internal space being conformally
balanced. These vacua are allowed to be warped along fiber or base
directions which are mediated by $[G]: {\bf 1}$ and $[H]: {\bf 3+\bar
3}$ components, respectively.


\subsubsection{Case (II)}


Now, we want to consider the second class of $SU(3)$ \N=1 Killing spinors,
where $a_L a_R \neq 0$. It turns out that all flux components can be non-zero
for this case. The different contractions of (\ref{380}) (see eqs.\
(\ref{383}) -- (\ref{386}) in the appendix) yield
\begin{eqnarray}
 e^{3U}\,{a_-^2 \over a_L a_R^\star}\, W&=&
\frac{1}{144}\,
G\haken (J\wedge J) +
i\,
 \frac{1}{36}\Big[ {a_L  \over a_R^\star} (H \haken \bar
\Psi)+ {a_R^\star \over a_L} (H \haken  \Psi) \Big] \label{501} \ , \\
\Big(1-{|a_L|^2 \over |a_R|^2}\Big)
 \, m &=&2\Big[{a_L \over a_R^\star} (H \haken \bar \Psi)+
{a_L^* \over a_R} (H \haken \Psi)\Big]
\label{504} \ , \\
e^{3U} \, \partial_a U&=& \frac{a_-^2 }{36\, a_+^2}
(H \haken (J \wedge J))_a
+ \frac{1}{144\, a_+^2}
 G\haken (J\wedge J) \; v_a   \label{505a}\ , \\
(H \haken J)_p
&=& \Big[ \frac{a_+^2}{12 \, a_L^\star a_R^\star}(G \haken \Psi)_p
+cc \Big]
\label{506}
\end{eqnarray}
Therefore, the superpotential and the Freund-Rubin parameter are fixed by the
singlet components of $G$ and $H$. Warp factor fixes  the Killing spinor
(\ref{523}) and (\ref{524}), ${\bf 3}+ {\bf \bar 3}$ components of $H$ and the
singlet component of $G$. Finally, the ${\bf 3}+ {\bf \bar 3}$ components of
$G$ are  fixed by the ${\bf 3 + \bar 3}$ components of $H$. Thus, the warp
factor $U$ and the $U(1)$ phase of the $SU(3)$ singlet spinor $\theta$ are not
fixed. The phase remains free, but the warp factor has to be fixed by the
equations of motion or Bianchi identity.  Since all fluxes can be non-zero,
the calculations of the torsion components becomes very involved and hence we
want to consider only specific examples.

\bigskip

\noindent
{\bf Example (1): \ $ \bf ([H]:3+\bar 3) + ([G]: 3 + \bar 3)$ }

\medskip

\noindent This case is equivalent to a Minkowski vacuum ($W=m=0$)
and the two vectors fix the warp factor. We should add a warning
here: due to the constraints (\ref{506}) that we used e.g.\ in
(\ref{505a}), there is no smooth limit to Case (I)! Since the
singlet of $G$ is zero, the warp factor depends only on the base
coordinates, ie.\ $v \haken \partial U = 0$.

The 4-form can be written as $e^{-3U}F= \frac{1}{2}(\Psi \wedge
\bar V +\bar \Psi\wedge  V) + v \wedge J \wedge V_o$ and therefore
the differential equations for the $G$-structures, when projected
on $X_6$ become (see eqs.\ (\ref{395}), (\ref{398}) and
(\ref{500}) in the appendix)
\begin{eqnarray}
(dv)_{pq}  & = & \frac{a_+^2e^{-3U}}{6a_-^2} F^{ cde}_{\ \ \
[p}(J\wedge J)_{q]cde} -\frac{ie^{-3U}}{a_-^2}\Big[a_La_R H^{
de}_{\ \ [p}\bar \Psi_{q]de})
 -cc \Big]  \label{546}
\\
(dJ)_{pqr} & = & \frac{6a_+^2e^{-3U}}{a_-^2} H^d_{\ [pq} J_{r]d}
 - \frac{3e^{-3U}}{a_-^2}\Big(a_La_R F^{de}_{\ \ [pq}\Psi_{r]de}+ cc
\Big)\nonumber \\&&
 - \partial_{[p}  \log (e^{-3U} a_-^2) J_{qr]} \label{547}
\\
 (d\Psi)_{pqrs} & = & -6e^{-3U} F^{fg}_{\ \ [pq}(v \wedge
\Psi)_{rs]fg}   -
\frac{4a_L^*a_R^*e^{-3U}}{a_-^2}\Big[\frac{1}{8}F^{efg}_{\ \ \
[p}(\Psi\wedge \bar\Psi)_{qrs]efg} \nonumber \\
&& +4 F^e_{\ [pqr}J_{s]e} -\frac{3i}{2}F^{ef}_{\ \ [pq}(J\wedge
J)_{rs]ef}+6iF_{pqrs}\Big] \nonumber \\
&& +  4\Big(7\partial_{[p} U + \frac{2}{a_-^2}(-a_L^*\partial_{[p}
a_L+ a_R^* \partial_{[p} a_R ) \Big) \Psi_{qrs]} \label{548}
\end{eqnarray}
When comparing with eqs.\ (\ref{191}-\ref{193}), one can find the following
 non-trivial torsion classes
\begin{eqnarray}
  \tau \in {\cal W}_4 \oplus {\cal W}_5  \ .
\end{eqnarray}
This implies that 6-d base conformal to a K\"ahler space.

\bigskip

\noindent
{\bf Example (2): \ $\bf m + ([H]:1+\bar 1)$ }

\medskip

If only Freund-Rubin parameter and $1+\bar 1$ components of $H$ are present,
we obtain an AdS vacua without warping, with the superpotential given by
\begin{eqnarray}
a_+^2 W&=& \frac{e^{-3U}}{36}[i a_L^2(H \haken \bar
\Psi)-ia_R^{*2}(H \haken  \Psi)] -a_La_R^*\frac{im}{36}e^{-3U}
\label{533}
\end{eqnarray}
and the internal flux reads $F = v \wedge H = {1 \over 2} e^{3U} v
\wedge (Q \Psi +cc)$.  In addition, it follows from (\ref{524}) that
$a_-^2$ does not depend on the base space coordinates.

Now, the differential equations of the G-structures, again projected on $X_6$,
become
\begin{eqnarray}
(dv)_{pq}  & = & 0
 \label{512}
\\
(dJ)_{pqr} & = & \frac{6a_+^2e^{-3U}}{a_-^2} H^d_{\ [pq} J_{r]d}
+\frac{3 }{a_-^2}\Big[(\bar W a_L^{2} +  W a_R^{2} )\Psi_{pqr}+ cc
\Big] \label{513}
\\
 (d\Psi)_{pqrs} & = & \frac{8}{a_-^2}(-a_L^*\partial_{[p} a_L+ a_R^*
\partial_{[p} a_R ) \Psi_{qrs]}
-\frac{4i }{a_-^2}( W a_L^{*2} + \bar W a_R^{*2} )(J\wedge
J)_{pqrs} \label{514}
\end{eqnarray}
When comparing with eqs.\ (\ref{191}-\ref{193}), one finds the following
non-trivial torsion classes:
\begin{eqnarray}
\tau \in  {\cal W}_1 \oplus {\cal W}_5
\end{eqnarray}
and hence the base space is not anymore a complex manifold. For an
appropriate choice of the phases of $a_L$ and $a_R$ this space
becomes nearly K\"ahler.

\bigskip

{\bf Example (3): \ $\bf [G]: 1$  }

\medskip

Finally, we want to consider the case where only the singlet
component of $G$ is non-zero, ie.\  the 4-form is now $F = e^{3U}
P\, J \wedge J$. This yields an AdS vacua with warping, but the
warp factor does not depend on coordinates of the base. Ie.\ we
have
\begin{eqnarray}
a_+^2 W&=& \frac{a_La_R^* }{144}e^{-3U}
G\haken (J\wedge J)\label{534} \\
a_+^2\partial U \haken v &=& \frac{a_-^2}{144}e^{-3U} G\haken
(J\wedge J) \label{535} \\
\partial_p U &=& 0 \label{536}
\end{eqnarray}
and again from (\ref{524}) we infer again that also $ \partial_p a_-^2 = 0$.
When projected on $X_6$, the $G$ structures become
\begin{eqnarray}
(dv)_{pq}  & = & \frac{a_+^2e^{-3U}}{6a_-^2} G^{ cde}_{\ \ \
[p}(J\wedge J)_{q]cde}
 \label{541}
\\
(dJ)_{pqr} & = & - \frac{3e^{-3U}}{a_-^2}\Big(a_La_R G^{de}_{\ \
[pq}\Psi_{r]de}+ a_L^*a_R^*G^{de}_{\ \ [pq}\bar \Psi_{r]de}\Big)
\nonumber \\&& +\frac{3 }{a_-^2}\Big((\bar W a_L^{2} +  W a_R^{2}
)\Psi_{pqr}+(W a_L^{*2} + \bar W a_R^{*2} )\bar \Psi_{pqr} \Big)
\label{542}
\\
 (d\Psi)_{pqrs} & = & \frac{4a_L^*a_R^*e^{-3U}}{3a_-^2}
\Big[\frac{9i}{2}G^{ef}_{\ \ [pq}(J\wedge
J)_{rs]ef}+6iG_{pqrs}\Big]\nonumber \\
&& + \frac{8}{a_-^2}(-a_L^*\partial_{[p} a_L+ a_R^* \partial_{[p}
a_R ) \Psi_{qrs]} -\frac{4i }{a_-^2}( W a_L^{*2} + \bar W a_R^{*2}
)(J\wedge J)_{pqrs} \label{543}
\end{eqnarray}
When comparing with eqs.\ (\ref{191}-\ref{193}), one can find the following
non-trivial torsion classes:
\begin{eqnarray}
  \tau \in  {\cal W}_1 \oplus {\cal W}_5
\end{eqnarray}
So, the geometry of the base space is the same as in the last
case, ie.\  an proper phase, it becomes a nearly K\"ahler space.
In fact, for any AdS compactification the 6-d base of the internal
space cannot be complex, see eq.\ (\ref{500}). A non-vanishing
superpotential necessarily leads to a non-vanishing torsion class
${\cal W}_1$, implying a non-complex base.

In summary, we have found flux vacua, satisfying Killing spinor constraints,
where all flux components can be turned on, with $W$, $\partial_m U$ and
$\partial_p U$ mediated by $\{m, [G]:1, [H]: 1+\bar 1\}$, $\{[G]: 1\}$ and
$\{[G]: 3+\bar 3, [H]:3+\bar 3\}$, respectively. Depending on which fluxes are
turned on, the base manifold of the internal space can either be K\"ahler
(with $W=0$) or nearly-K\"ahler (with $W \neq 0$).  Note, we have presented
only the necessary conditions; there may be additional constraints, in
particular, arising from the equations of motion and Bianchi identities.


\subsection{$G_2$ Structures}


This case is related to the simplest spinor Ansatz (\ref{N=1(II)}) and
can be obtained from the general $SU(3)$ spinor by imposing $a_L =
a_R^\star$. This yields the following two equations
\begin{eqnarray}
 0&=&  W a \theta+ (\frac{1}{2}
\partial U + \frac{im}{72}e^{-3U} + \frac{1}{144}e^{-3U}  F ) \,
a \theta \label{303} \\
\nabla_a \theta &=&y  \gamma_a  W
\theta^*+(\frac{e^{-3U}}{12} F_a+ \frac{1}{2}\partial_a U-
\frac{\partial_a a}{a} +\frac{ime^{-3U}\gamma_a}{48})\theta
\label{304}
\end{eqnarray}
Combining these two equations and their complex conjugate, we find
\begin{eqnarray}
  W_1 \theta + \frac{1}{144}e^{-3U}  F  \,
\theta=0  \label{241}\\
   i W_2 \theta + (\frac{1}{2}
\partial U + \frac{im}{72}e^{-3U} ) \,
\theta =0 \label{242} \\
 \gamma_a  W_1 \theta+\frac{e^{-3U}}{12}
F_a \theta=0 \label{243}\\
 i \gamma_a W_2 \theta - [\nabla_a
-(\frac{1}{2}\gamma_a\partial
U+\frac{ime^{-3U}\gamma_a}{48}-\frac{\partial_a a}{a})]\theta=0
\label{244}
\end{eqnarray}
which are just the $G_2$ \N = 1 Killing spinor equations which are
discussed in \cite{0311119}.

\medskip

Since we obtained no new results,  let us only summarize  the results.
Contracting eq.\ (\ref{243}) with $\gamma^a$ and then comparing it with eq.\
(\ref{241}), we find that no internal fluxes can be turned on. Contracting
eq.\ (\ref{242}) with $\theta^T\gamma_{ab}$, we have
\begin{eqnarray}
\partial_a U=0
\end{eqnarray}
and hence (set $dU=0$)
\begin{eqnarray}
 W_2=\frac{-m}{72} \ .
\end{eqnarray}
Then eq.\ (\ref{244}) leads to
\begin{eqnarray}
\nabla_a \theta =\Big[\frac{im\gamma_a}{36}-\frac{\partial_a
a}{a}\Big]\theta \label{245}
\end{eqnarray}
On the other hand, $G_2$-structures require
\begin{eqnarray}
\nabla \Sigma^{(0)}=\nabla \Sigma^{(1)}=\nabla \Sigma^{(2)}=0
\end{eqnarray}
The first condition implies $da =0$ and the other two are
automatically satisfied. The differential equations for the
$G$-structure are now simply
\begin{eqnarray}
d\varphi =\frac{2m}{9} \psi, \ \ \ \ d \psi=0
\end{eqnarray}
implying that the internal space admits weak $G_2$ holonomy; for more
details for these vacua we refer to \cite{0111274}.

Finally, it is obvious that Bianchi identity and equations of motion can be
satisfied. To conclude, this case does not allow for internal fluxes;  the
internal space has weak $G_2$ holonomy and its cosmological constant is given
by the Freund-Rubin parameter.


\section{Discussions and Conclusions}
\resetcounter


In this paper we presented a systematic classification of
supersymmetric vacua from compactifications of M-theory on a
general seven-dimensional manifold in the presence of general
four-form fluxes. Any seven-dimensional spin manifold admits three
globally well-defined vectors and with these vectors one can
always define $SU(2)$ structures (which includes the cases with
$SU(3)$ and $G_2$ structures).  At the same time, these vectors
imply a fibration of the seven-dimensional manifold over a
four-dimensional base $X_4$ for $SU(2)$ and over a six-dimensional
base $X_6$ for $SU(3)$ structures. We will now summarize which
flux components can be non-zero and what is the resulting geometry
of $X_4$ and $X_6$.

Depending on the number of external spinors, the vacua have \N =
1, \N= 2, \N=3 or \N = 4 supersymmetry in four dimensions. For the
\N = 4 case eq. (\ref{N=4}), no fluxes can be turned on (while
preserving at least $SU(2)$ structures). We did not discuss in
detail the \N = 3 case; we give the relevant equations in Section
4, which need however further exploration. But we discussed in
detail the cases with \N = 2 and \N = 1 supersymmetry and
summarized them in two tables.

\begin{table}[h]
\caption{Non-trivial flux components and their effects on
mass matrix $W$ and warp factor $U$ in all \N = 2 cases are
summarized. Constraints from Bianchi identities and equations of
motion are not included in this table. We turned off non-regular
fluxes, ie.\  the B and D parts, which can mediate $\partial_p
U$. For further notations see Section 3.4.\label{1000}} \vspace{0.4cm}
\begin{center}
\begin{tabular}{|c|c|c|c|c|}
\hline \N = 2 &\multicolumn{3}{|c|}{$SU(2)$-Structures}& $SU(3)$\\
\cline{2-4}  & Case (a) & Case (b) & Case (c) & -Structures \\
\hline
  W & / & / & / & $m$
  \\
\hline
 $\partial_m U$& $[C]: {\bf 1}$ &$[A]$ & / & /
  \\
\hline
 $\partial_p U$& / & /  &
/ & /
  \\
\hline {\rm Allowed Fluxes} & $[C]: {\bf 1+3}$ & $[A]$ & / & $m$
  \\
\hline
\end{tabular}
\end{center}
\end{table}
There are two classes of \N = 2 vacua, one with $SU(2)$ structures
and another one with $SU(3)$ structures. For the $SU(3)$ case, all
internal fluxes have to be trivial and only the Freund-Rubin
parameter can be non zero and the external space is AdS and the
internal space is Einstein-Sasaki. For the vacuua admitting
$SU(2)$ structures, we consider three cases related to different
chiral choice of the two external spinors, eg.\ whether both
spinors have the same chirality, opposite or whether one is a
Majorana spinor, see spinor Ans\"atze in eqs.\ (\ref{N=2(I)a}),
(\ref{N=2(I)b}) and (\ref{N=2(I)c}). Note, in all cases $W$ has to
vanish and therefore the external space is flat. The
four-dimensional base manifold of the internal space is conformal
to a K\"ahler space.

\begin{table}[h]
\caption{Non-trivial flux components and their effects on
superpotential and warped factor in all \N = 1 cases are
summarized. Constraints from Bianchi identities and equations of
motion are not included in this table. \label{1001}}
\vspace{0.4cm}
\begin{center}
\begin{tabular}{|c|c|c|c|}
\hline \N = 1 &\multicolumn{2}{|c|}{$SU(3)$-Structures}& $G_2$\\
\cline{2-3}  &  Case (I) ($a_R=0$) &  Case (II) ($a_L a_R \neq 0$)
&
-Structures\\
\hline
  W & / & $m$, $[G]: {\bf 1}$, $[H]: {\bf 1+\bar 1}$ & $m$
  \\
\hline
 $\partial_m U$& $[G]:{\bf 1}$ &$[G]: {\bf 1}$ & /
  \\
\cline{1-3}
 $\partial_p U$& $[H]: {\bf 3+ \bar 3}$ & $[G]:{\bf  3+\bar 3}$,
$[H]: {\bf 3+\bar 3}$ &
  \\
\hline {\rm Allowed Fluxes} & $[G]: {\bf 8+1}$ & $m$, $[G]$, $[H]$ & $m$\\
&  $[H]: {\bf 6+\bar 6+3+\bar 3}$ &&
  \\
\hline
\end{tabular}
\end{center}
\end{table}
For \N = 1 vacua we found also two classes, one with $SU(3)$
structures and one with $G_2$ structures. The latter case is very
similar to the \N = 2 vacuum with $SU(3)$ structures. Here, only
the Freund-Rubin parameter can be non-zero, there is no warping
and the internal space has weak $G_2$ holonomy, ie.\ it is an
Einstein space. The external space is AdS. Non-trivial internal
fluxes are only allowed if the structure group is only $SU(3)$ and
the two cases are again related to different chiral choices; see
spinor Ansatz in eq.\ (\ref{N=1(I)}). Case (I) corresponds to the
case, which has been discussed already in the literature. Some
flux components are not allowed and the superpotential has to
vanish and thus the external space is flat. The six-dimensional
base of the internal space is conformal to a balanced manifold. On
the other hand, Case (II) has not  been discussed in the
literature. Here, all fluxes and the superpotential can be
non-zero. We discussed special cases where the base of the
internal space becomes conformal to a K\"ahler space (by setting
$W=0$) or it can be nearly K\"ahler, which requires $W \neq 0$. In
general, an AdS vacuum requires that the base of the internal
space is non-complex. As we pointed out before, Killing spinor in
case (I) has definite 10D chirality (since $\theta$ is 6 dim
chiral spinor) and can survive $S_1/Z_2$ projection with $S_1$
being $SU(3)$ fiber in the internal space. So it can be embedded
in heterotic M theory. And the Killing spinor in case (II) only
admits $S^1$ fiberation of $SU(3)$ structures and hence matches
with M theory lift of pure type IIA supergravity.

There are a number of directions  that are interesting for future
exploration. We  did not analyzed the \N = 3  in detail and it
would be interesting to work out detailed  constraints on the
fluxes as well as the geometry in this case. In addition we did
not consider new examples with explicit solutions for the metric
and flux components.  It would be interesting, at least for the \N
= 1 case,  to work out some new explicit solutions, that solve the
Killing spinor constraints, and to further  investigate the
constraints imposed by the  Bianchi identities and the equations
of motion for fluxes. Finally, it would be interesting to explore
further  the relation of these supersymmetric vacua  to the known
flux vacua of  ten-dimensional  Type IIA string theory, and also
the vacua of Type IIB  string theory, as eg.\  the explicit
vacua found \cite{0412250, 0502154}.

\bigskip

\bigskip

\noindent{\Large{\bf Acknowledgements}} We would like to thank
Peng Gao and Claus Jeschek for useful discussions.  Research is supported in
part by the Department of Energy grant DE-FG03-95ER40917 (M.C. and
T.L.), National Science Foundation grant INTO3-24081 (M.C. and
T.L.), the University of Pennsylvania Research Foundation Award
(M.C.), and the Fay R. and Eugene L. Langberg Chair (M.C.).



\appendix

\section*{Appendix}

\resetcounter

\renewcommand{\theequation}{A.\arabic{equation}}


\section{Notation and Conventions}


Here we summarize  our notation and conventions.  The flat $\Gamma$ matrix
algebra reads $ \{ \Gamma^A , \Gamma^B \} = 2 \eta^{AB}$ with $\eta = {\rm
diag}(-,+,+ \ldots +)$, we decompose the $\Gamma$-matrices as usual
\be391 \Gamma^\mu = \hat \gamma^\mu \otimes \mathbb{I} \qquad ,
\qquad \Gamma^{a+3} = \hat \gamma^5 \otimes \gamma^a \ee
with $\mu = 0,1,2,3$, $a = 1,2, \ldots 7$ and
\be791
\hat \gamma^5 = i \hat \gamma^0 \hat \gamma^1 \hat \gamma^2 \hat
\gamma^3 \ , \quad  \gamma^1  \gamma^2  \gamma^3 \gamma^4 \gamma^5
         \gamma^6  \gamma^{7} = - i
\ee
which implies
\be161 i \hat \gamma^5 \hat \gamma^\mu = {1 \over 3!}
\varepsilon^{\mu\nu\rho\lambda} \hat \gamma_{\nu\rho\lambda} \quad
,\qquad {i \over 3!} \varepsilon^{abcdmnp} \gamma_{mnp} =
\gamma^{abcd} \equiv \gamma^{[a} \gamma^b \gamma^c \gamma^{d]} \ .
\ee
With the identity
\begin{eqnarray}
\Gamma_M \Gamma_{N_1 \cdots N_n} = \Gamma_{M N_1 \cdots N_n} +
n \, G_{ M [N_1} \Gamma_{N_2 \cdots N_n]} \nonumber \\
\Gamma_{N_1 \cdots N_n} \Gamma_M = \Gamma_{ N_1 \cdots N_n M} + n
\,  \Gamma_{[N_1 \cdots N_{n-1}}\delta_{N_n] M} \label{111}
\end{eqnarray}
the decomposition implies
\begin{eqnarray}
[\gamma_a,\gamma_{b_1\cdots b_n}] = \left\{ \ba{ll}
 2\gamma_{ab_1\cdots b_n}, \ & {\rm \ n \ is \ odd} \\
 2n\delta_{a[b_1}\gamma_{b_2\cdots b_n]} \ & {\rm n \ is \ even} \ea
 \right.  \label{01}
\end{eqnarray}
and
\begin{eqnarray}
\{\gamma_a,\gamma_{b_1\cdots b_n}\} = \left\{ \ba{ll}
 2n\delta_{a[b_1}\gamma_{b_2\cdots b_n]} , \ & {\rm n \ is \ odd} \\
 2\gamma_{ab_1\cdots b_n}\ & {\rm n \ is \ even} \ea
 \right. \label{02}
\end{eqnarray}
both of which are very useful for our future purpose. The spinors
in 11-d supergravity are in the Majorana representation and hence,
all 4-d $\hat \gamma^\mu$-matrices are real and $\hat \gamma^5$ as
well as the 7-d $\gamma^a$-matrices are purely imaginary and
antisymmetric.

A $G_2$ singlet spinor obeys the following
multiplication with $\gamma$-matrices
\be912
\ba{rcl}
\gamma_{abc} \theta_0 &=& \Big( i \varphi_{abc} +
\psi_{abcd} \gamma^d \Big)
  \, \theta_0 \ , \\
\gamma_{abcd} \theta_0 &=& \Big( - \psi_{abcd} - 4 i
\varphi_{[abc}
        \gamma_{d]} \Big) \theta_0 \ .
\ea \ee

The Index conventions are as follows: ``p--t'' denoting the base
directions, ``i--n'' denoting the fiber directions, and ``a--g''
denoting the whole 7-d internal space.

The contraction ``$\haken$'' is defined as
\[
A\haken
B= B\haken A = A^{p_1...p_m} B_{p_1...p_mp_{m+1}...p_n}
\]


\section{$G$ Structure Equations for \N = 2 Cases}

\resetcounter

\renewcommand{\theequation}{B.\arabic{equation}}


In this appendix we summarize the differential equations for the forms in the
case of  \N = 2 vacua.

\bigskip
\noindent
{\bf For Case (a)} we find
\begin{eqnarray}
\nabla_a \Sigma^{(k)}_{xy} & = & \theta^\dagger_x \Big
[\gamma^{(k)},\frac{e^{-3U}}{12} F_a+\frac{1}{2}\gamma_{ab} \partial^b
U\Big ] \theta_y + \partial_a [U - \log (a_x^\star a_y) ] \,
 \Sigma^{(k)}_{xy} \nonumber \\ && +
  \Big(\, \frac{a_{x}}{a_x^*}\bar{W}_x^z\theta_z^T\gamma_a\gamma^{(k)}
\theta_y+\frac{a_y^*}{a_y}\theta_x^{\dagger}\gamma^{(k)}\gamma_a
W_y^z\theta_z^* \, \Big) \label{G-structure equation 1} \ , \\
\nabla_a \Omega^{(k)}_{xy} & = & \theta^T_x \Big
\{\gamma^{(k)},\frac{e^{-3U}}{12} F_a\Big \} \theta_y + \theta^T_x
\Big [\gamma^{(k)}, \frac{1}{2}\gamma_{ab} \partial^b U \Big ]
\theta_y+ \partial_a [U - \log (a_x a_y) ] \, \Omega^{(k)}_{xy}
\nonumber \\ &&
- \Big(\, \frac{a_x^*}{a_x}W_x^z
\theta_z^\dagger\gamma_a\gamma^{(k)}\theta_y - \frac{a_y^*}
{a_y}\theta_x^T\gamma^{(k)}\gamma_a
W_y^z\theta_z^*\, \Big) \label{G-structure equation 2}\ .
\end{eqnarray}

\bigskip
\noindent
{\bf For Case (b)} we obtain
\begin{eqnarray}
\nabla_a  \Sigma^{(k)}_{11} & = & \theta^\dagger_1 \Big
[\gamma^{(k)},\frac{e^{-3U}}{12}
F_a+\frac{im\gamma_a }{48}+\frac{1}{2}\gamma_{ab}\partial^b
U\Big ]
 \theta_1 - \partial_a \log (e^{-U} |a_1|^2) \;  \Sigma_{11}^{(k)}
\nonumber \\
&& +
 \Big(\frac{a_2^*}{a_1^*}\bar{W}_{12}\theta_2^\dagger\gamma_a\gamma^{(k)}
\theta_1+\frac{a_2}{a_1}W_{12}\theta_1^{\dagger}\gamma^{(k)}\gamma_a
\theta_2 \nonumber
\\&& +\frac{a_1}{a_1^*}\bar{W}_{11}\theta_1^T \gamma_a\gamma^{(k)}\theta_1+
\frac{a_1^*}{a_1}W_{11}\theta_1^{\dagger}\gamma^{(k)}\gamma_a
\theta_1^*\Big)  \label{G-d equation1} \\
\nabla_a  \Sigma^{(k)}_{22} & = & \theta^\dagger_2 \Big
[\gamma^{(k)},\frac{-e^{-3U}}{12}
F_a+\frac{im\gamma_a }{48}+\frac{1}{2}\gamma_{ab}\partial^b
U\Big ] \theta_2 - \partial_a \log (e^{-U} |a_2|^2) \;  \Sigma_{22}^{(k)}
\nonumber \\
&& -
 \Big(\frac{a_1^*}{a_2^*}W_{21}\theta_1^\dagger\gamma_a\gamma^{(k)}
\theta_2+\frac{a_1}{a_2}\bar
W_{21}\theta_2^{\dagger}\gamma^{(k)}\gamma_a \theta_1\nonumber
\\&& +\frac{a_2}{a_2^*}W_{22}\theta_2^T \gamma_a\gamma^{(k)}\theta_2+
\frac{a_2^*}{a_2}\bar W_{22}\theta_2^{\dagger}\gamma^{(k)}\gamma_a
\theta_2^*\Big)  \label{G-d equation2} \\
\nabla_a  \Sigma^{(k)}_{12} & = & -\theta_1^\dagger
\Big\{\gamma^{(k)}, \frac{e^{-3U}}{12}
F_a\Big\}\theta_2+\theta^\dagger_1 \Big
[\gamma^{(k)},\frac{im\gamma_a }{48}+\frac{1}{2}\gamma_{ab}\partial^b
U\Big ]
 \theta_2 \nonumber \\
&& - \partial_a \log (e^{-U} a_1^\star a_2 )  \Sigma_{12}^{(k)}
+
 \Big(\frac{a_2^*}{a_1^*}\bar{W}_{12}\theta_2^\dagger\gamma_a\gamma^{(k)}
\theta_2-\frac{a_1}{a_2}\bar
W_{21}\theta_1^{\dagger}\gamma^{(k)}\gamma_a \theta_1 \nonumber
\\&& +\frac{a_1}{a_1^*}\bar{W}_{11}\theta_1^T \gamma_a\gamma^{(k)}\theta_2-
\frac{a_2^*}{a_2}\bar W_{22}\theta_1^{\dagger}\gamma^{(k)}\gamma_a
\theta_2^*\Big) \label{G-d
equation3}\\
\nabla_a  \Sigma^{(k)}_{21} & = & \theta_2^\dagger
\Big\{\gamma^{(k)}, \frac{e^{-3U}}{12}
F_a\Big\}\theta_1+\theta^\dagger_2 \Big
[\gamma^{(k)},\frac{im\gamma_a}{48}+\frac{1}{2}\gamma_{ab}\partial^b
U\Big ]
 \theta_1  \nonumber \\
&&
- \partial_a \log (e^{-U} a_2^\star a_1 )  \Sigma_{21}^{(k)}
 +
 \Big(-\frac{a_1^*}{a_2^*}W_{21}\theta_1^\dagger\gamma_a\gamma^{(k)}
\theta_1+\frac{a_2}{a_1}
W_{12}\theta_2^{\dagger}\gamma^{(k)}\gamma_a \theta_2\nonumber
\\&& -\frac{a_2}{a_2^*}W_{22}\theta_2^T
\gamma_a\gamma^{(k)}\theta_1+\frac{a_1^*}{a_1}W_{11}\theta_2^{\dagger}
\gamma^{(k)}\gamma_a
\theta_1^*\Big)  \label{G-d equation4}
\end{eqnarray}
and
\begin{eqnarray}
\nabla_a  \Omega^{(k)}_{11} & = & \theta_1^T
\Big\{\gamma^{(k)}, \frac{e^{-3U}}{12}
F_a\Big\}\theta_1+\theta^T_1 \Big
[\gamma^{(k)},\frac{ime^{-3U}\gamma_a}{48}
+\frac{1}{2}\gamma_{ab}\partial^b U \Big ]
 \theta_1   \nonumber \\
&& - \partial_a \log (e^{-U} a_1^2 ) \; \Omega_{11}^{(k)}
  + \Big(-\frac{a_2}{a_1}W_{12}\theta_2^T\gamma_a\gamma^{(k)}\theta_1+
\frac{a_2}{a_1}W_{12}\theta_1^T\gamma^{(k)}\gamma_a
\theta_2\nonumber
\\&& -\frac{a_1^*}{a_1}W_{11}\theta_1^\dagger \gamma_a\gamma^{(k)}\theta_1+
\frac{a_1^*}{a_1}W_{11}\theta_1^T\gamma^{(k)}\gamma_a
\theta_1^*\Big)  \label{G-d equation5} \\
\nabla_a  \Omega^{(k)}_{22} & = & \theta_2^T
\Big\{\gamma^{(k)}, \frac{-e^{-3U}}{12}
F_a\Big\}\theta_2+\theta^T_2 \Big
[\gamma^{(k)},\frac{ime^{-3U}\gamma_a}{48}
+\frac{1}{2}\gamma_{ab}\partial^b U\Big ]
 \theta_2  \nonumber \\
&& - \partial_a \log (e^{-U} a_2^2 ) \; \Omega_{22}^{(k)}
 +  \Big(\frac{a_1}{a_2}\bar
W_{21}\theta_1^T\gamma_a\gamma^{(k)}\theta_2-\frac{a_1}{a_2}\bar
W_{21}\theta_2^T\gamma^{(k)}\gamma_a \theta_1\nonumber
\\&& +\frac{a_2^*}{a_2}\bar W_{22}\theta_2^\dagger \gamma_a\gamma^{(k)}
\theta_2-\frac{a_2^*}{a_2}\bar W_{22}\theta_2^T\gamma^{(k)}\gamma_a
\theta_2^*\Big)  \label{G-d equation6} \\
\nabla_a  \Omega^{(k)}_{12} & = & \theta^T_1 \Big
[\gamma^{(k)},\frac{-e^{-3U}}{12} F_a+\frac{ime^{-3U}\gamma_a}{48}
+\frac{1}{2}\gamma_{ab}\partial^b U\Big ]
 \theta_2  \nonumber \\
&& - \partial_a \log (e^{-U} a_1 a_2 ) \; \Omega_{12}^{(k)}
- \Big(\frac{a_2}{a_1}W_{12}\theta_2^T\gamma_a\gamma^{(k)}\theta_2+
\frac{a_1}{a_2}\bar
W_{21}\theta_1^T\gamma^{(k)}\gamma_a \theta_1\nonumber
\\&& +\frac{a_1^*}{a_1}W_{11}\theta_1^\dagger
\gamma_a\gamma^{(k)}\theta_2+\frac{a_2^*}{a_2}\bar
W_{22}\theta_1^T\gamma^{(k)}\gamma_a \theta_2^*\Big)  \label{G-d
equation7}\\
\nabla_a \Omega^{(k)}_{21} & = & \theta^T_2 \Big
[\gamma^{(k)},\frac{e^{-3U}}{12} F_a+\frac{ime^{-3U}\gamma_a}{48}
+\frac{1}{2}\gamma_{ab}\partial^b U \Big ]
 \theta_1  \nonumber \\
&& - \partial_a \log (e^{-U} a_1 a_2 ) \; \Omega_{21}^{(k)}
+  \Big(\frac{a_1}{a_2}\bar
W_{21}\theta_1^T\gamma_a\gamma^{(k)}\theta_1+\frac{a_2}{a_1}
W_{12}\theta_2^T\gamma^{(k)}\gamma_a \theta_2\nonumber
\\&& +\frac{a_2^*}{a_2}\bar W_{22}\theta_2^\dagger
\gamma_a\gamma^{(k)}\theta_1+\frac{a_1^*}{a_1}W_{11}\theta_2^T\gamma^{(k)}
\gamma_a \theta_1^*\Big)  \label{G-d equation8}
\end{eqnarray}

\bigskip
\noindent
{\bf For Case (c)} we obtain
\begin{eqnarray}
\nabla_a \hat \Sigma^{(k)}_{xy} & = & \hat \theta^\dagger_x \Big
[\gamma^{(k)},\frac{e^{-3U}}{12} F_a+\frac{1}{2}\gamma_{ab}
\partial^b U\Big ]
 \hat \theta_y
- \partial_a \log (e^{-U} a_x^\star a_y ) \; \hat \Sigma_{xy}^{(k)}
\nonumber \\
&& +  (\frac{a_{x}}{a_x^*}\bar{W}_x^z\hat
\theta_z^T\gamma_a\gamma^{(k)}\hat \theta_y+\frac{a_y^*}{a_y}\hat
\theta_x^{\dagger}\gamma^{(k)}\gamma_a W_y^z\hat \theta_z^*)
\label{421}
\\
\nabla_a   {\hat \Omega}^{(k)}_{xy} & = & \hat \theta^T_x
\Big \{\gamma^{(k)},\frac{e^{-3U}}{12} F_a\Big \} \hat \theta_y  +
\hat \theta^T_x \Big [\gamma^{(k)}, \frac{1}{2}\gamma_{ab}
\partial^b U \Big ]
 \hat \theta_y
- \partial_a \log (e^{-U} a_x a_y ) \; \hat \Omega_{xy}^{(k)}
\nonumber \\ &&
-  (\frac{a_x^*}{a_x}W_x^z \hat
\theta_z^\dagger\gamma_a\gamma^{(k)}\hat
\theta_y - \frac{a_y^*}{a_y}\hat \theta_x^T\gamma^{(k)}\gamma_a
W_y^z\hat \theta_z^*) \label{422}
\end{eqnarray}


\section{More Results for $SU(3)$ ${\cal N}=1$ Case}


According to the definition of $\chi$ eq.\ (\ref{382}),
we have
\begin{eqnarray}
\chi^{\dagger} \gamma^{(k)} \chi &=& |a_L|^2 \Sigma^{(k)} +
(-1)^k|a_R|^2 \Sigma^{(k)*}+a_La_R \Omega^{(k)} +
a_La_R^*(-1)^k\Omega^{(k)*}  \label{519}\\
\chi^T \gamma^{(k)} \chi &=& a_L^2 \Omega^{(k)} + a_R^{*2}
\Omega^{(k)*}+a_La_R^* [\Omega^{(k)} + (-1)^k\Omega^{(k)*}]
\label{520}
\end{eqnarray}
with
\begin{eqnarray}
\chi^T \gamma_a \chi = \chi^T \gamma_{ab} \chi = 0 \label{521}
\end{eqnarray}

Let us present some more useful identities.  Contracting eq.\
(\ref{380}) with $\theta^T$ and $\theta^{\dagger}$, we have
\begin{eqnarray}
a_L^* W&=& \frac{ia_L e^{-3U}}{36}(H \haken \bar \Psi) \nonumber
\\&& - a_R^*\Big(\frac{-1}{2} \partial U \haken v +
\frac{im}{72}e^{-3U} - \frac{1}{288}e^{-3U} G\haken (J\wedge J)
\Big)\label{383}\\
a_RW&=& \frac{-ia_R^* e^{-3U}}{36}(H \haken \Psi) \nonumber \\&& -
a_L\Big(\frac{1}{2} \partial U \haken v + \frac{im}{72}e^{-3U} -
\frac{1}{288}e^{-3U} G\haken (J\wedge J) \Big)\label{384}
\end{eqnarray}
and contracting eq.\ (\ref{380}) with $\theta^T\gamma_p$ and
$\theta^{\dagger}\gamma_p$, we have
\begin{eqnarray}
0&=& \frac{-ia_Le^{-3U}}{36}(G \haken \Psi)_p
+a_R^*\Big(\frac{1}{2}\partial_p U+\frac{i}{2} (\partial U\haken
J)_p \nonumber
\\&& +\frac{e^{-3U}}{72}(H\haken(J\wedge J))_p -
\frac{ie^{-3U}}{12} (H\haken J)_p\Big) \label{385}
\\
0&=& \frac{-ia_R^*e^{-3U}}{36}(G \haken \bar \Psi)_p
+a_L\Big(\frac{1}{2}\partial_p U-\frac{i}{2} (\partial U\haken
J)_p \nonumber
\\&& -\frac{e^{-3U}}{72}(H\haken(J\wedge J))_p -
\frac{ie^{-3U}}{12} (H\haken J)_p\Big) \label{386}
\end{eqnarray}

The internal Killing spinor equation eq.\ (\ref{381})
yields the differential equations for the $G$-structure
\begin{eqnarray}
\nabla_a  \Sigma^{(k)} & = & \theta^\dagger \Big
[\gamma^{(k)},\frac{a_+^2e^{-3U}}{12a_-^2} F_a +
\frac{1}{2}\gamma_{ab}
\partial^b U +\frac{ime^{-3U}\gamma_a}{48}\Big ]
 \theta \nonumber \\&&
 + \frac{e^{-3U}}{6a_-^2}[-a_La_R\theta^T F_a \gamma^{(k)}\theta+
a_L^*a_R^*\theta^\dagger  \gamma^{(k)}F_a\theta^*]\nonumber \\
&& + \Big(\partial_a U - \partial_a \log a_-^2 \Big) \theta^\dagger
\gamma^{(k)}
\theta \nonumber \\
&& + \frac{1}{a_-^2}(-a_L\partial_a a_R+a_R\partial_a a_L)\theta^T
\gamma^{(k)} \theta + \frac{1}{a_-^2}(-a_L^*\partial_a
a_R^*+a_R^*\partial_a a_L^*)\theta^\dagger \gamma^{(k)} \theta^*
\nonumber \\
&&+ \frac{1 }{a_-^2}(W a_L^*a_R + \bar W a_L a_R^*)\theta^\dagger
\{\gamma_a, \gamma^{(k)} \}\theta \nonumber \\
&& +\frac{1 }{a_-^2}\Big((\bar W a_L^{2} +  W a_R^{2}
)\theta^T\gamma_a\gamma^{(k)}\theta+(W a_L^{*2} + \bar W a_R^{*2}
)\theta^{\dagger}\gamma^{(k)}\gamma_a \theta^* \Big) \label{388}
\end{eqnarray}
and
\begin{eqnarray}
\nabla_a \Omega^{(k)} & = & \theta^T \Big
\{\gamma^{(k)},\frac{a_+^2e^{-3U}}{12a_-^2} F_a\Big\}\theta+
\theta^T \Big [\gamma^{(k)}, \frac{1}{2}\gamma_{ab} \partial^b U
+\frac{ime^{-3U}\gamma_a}{48}\Big ] \theta \nonumber \\
&& + \frac{a_L^*a_R^*e^{-3U}}{6a_-^2}[\theta^\dagger F_a
\gamma^{(k)}\theta+ \theta^T \gamma^{(k)}F_a\theta^*]\nonumber \\
&& + \Big(\partial_a U + \frac{2}{a_-^2}(-a_L^*\partial_a a_L+
a_R^*
\partial_a a_R ) \Big) \theta^T \gamma^{(k)} \theta \nonumber \\
&& + \frac{1}{a_-^2}(-a_L^*\partial_a a_R^*+a_R^*\partial_a
a_L^*)(\theta^\dagger \gamma^{(k)} \theta + \theta^T \gamma^{(k)}
\theta^*) \nonumber \\&&+ \frac{1 }{a_-^2}(W a_L^*a_R + \bar W
a_L a_R^*)\theta^T[ \gamma^{(k)}, \gamma_a]\theta \nonumber \\
&& +\frac{1 }{a_-^2}( W a_L^{*2} + \bar W a_R^{*2}
)(-\theta^\dagger
\gamma_a\gamma^{(k)}\theta+\theta^T\gamma^{(k)}\gamma_a \theta^* )
\label{389}
\end{eqnarray}
The differential equation of the global vector field ($\sim \Sigma^{(1)}$)
is therefore
\begin{eqnarray}
\nabla_a  v_b & = & \frac{a_+^2e^{-3U}}{12a_-^2} F_a^{\
cde}(J\wedge J)_{cdeb} + 2\delta_{b[a}v_{c]}\partial^c U
+\frac{ime^{-3U}\delta_{ab}}{24}
 \nonumber \\&&
 + \frac{ie^{-3U}}{6a_-^2}\Big[a_La_R((\bar \Psi\haken F)_a v_b -
  3H_a^{\ de}\bar \Psi_{deb})
 - a_L^*a_R^*((\Psi\haken F)_a v_b -3H_a^{\ de}\Psi_{deb})\Big]\nonumber \\&&
 +
 \Big(\partial_a U - \partial_a \log a_-^2  \Big) v_b  +
\frac{2 }{a_-^2}(W a_L^*a_R + \bar W a_L a_R^*)\delta_{ab}
\label{394}
\end{eqnarray}
where the second term interprets the conformal scaling effect and
the other terms come from $\tilde \nabla_a v_b$ with respect to
$h_{ab}$. Antisymmetrizing this equation, we obtain
\begin{eqnarray}
(dv)_{ab}  & = & \frac{a_+^2e^{-3U}}{6a_-^2} F^{ cde}_{\ \ \
[a}(J\wedge J)_{b]cde}  -
 2v_{[a} \partial_{b]}(2 U - \log a_-^2)
 \nonumber \\&&
 + \frac{ie^{-3U}}{3a_-^2}\Big[a_La_R((\bar \Psi\haken F)_{[a} v_{b]}
-3H^{ de}_{\ \ [a}\bar \Psi_{b]de})
 \nonumber \\&&
 - a_L^*a_R^*((\Psi\haken F)_{[a} v_{b]} -3H^{ de}_{\ \ [a}\Psi_{b]de})\Big]
 \label{395}
\end{eqnarray}
For almost complex structure $J$, we have
\begin{eqnarray}
\nabla_a  J_{bc} & = & \frac{-a_+^2e^{-3U}}{a_-^2} F^{ef}_{\ \
a[b}(v \wedge J)_{c]ef})-2
\partial^d U (\delta_{a[b}J_{c]d}-\delta_{d[b}J_{c]a}) \nonumber
\\&&
 - \frac{e^{-3U}}{a_-^2}\Big(a_La_R F^{ef}_{\ \ a[b}\Psi_{c]ef}+
a_L^*a_R^*F^{ef}_{\ \ a[b}\bar \Psi_{c]ef}\Big)\nonumber \\&&
 +
 \Big(\partial_a U - \partial_a \log a_-^2 \Big) J_{bc}  +
 \frac{2 }{a_-^2}(W a_L^*a_R + \bar W a_L
a_R^*)(v\wedge J)_{abc} \nonumber
\\&&
+\frac{1 }{a_-^2}\Big((\bar W a_L^{2} +  W a_R^{2} )\Psi_{abc}+(W
a_L^{*2} + \bar W a_R^{*2} )\bar \Psi_{abc} \Big) \label{396}
\end{eqnarray}
and hence
\begin{eqnarray}
(dJ)_{abc} & = & \frac{a_+^2e^{-3U}}{a_-^2} \Big(-3(J \haken
G)_{[ab}v_{c]}+6H^d_{\ [ab} J_{c]d}\Big)  \nonumber \\&&
 - \frac{3e^{-3U}}{a_-^2}\Big(a_La_R F^{de}_{\ \ [ab}\Psi_{c]de}+
a_L^*a_R^*F^{de}_{\ \ [ab}\bar \Psi_{c]de}\Big)\nonumber \\&&
 +
3 \partial_{[a}( 3 U -  \log a_-^2 ) J_{bc]}  + \frac{6 }{a_-^2}(W
a_L^*a_R + \bar W a_L a_R^*)(v\wedge J)_{abc} \nonumber
\\&&
+\frac{3  }{a_-^2}\Big((\bar W a_L^{2} +  W a_R^{2} )\Psi_{abc}+(W
a_L^{*2} + \bar W a_R^{*2} )\bar \Psi_{abc} \Big) \label{398}
\end{eqnarray}
Finally, let's consider the differential equation of the invariant
three form $\Psi$.
\begin{eqnarray}
\nabla_a \Psi_{bcd} & = & \frac{-3e^{-3U}}{2} F^{fg}_{\ \ a[b}(v
\wedge \Psi)_{cd]fg} +\frac{ime^{-3U}}{24}(v\wedge \Psi)_{abcd} +
6 \partial^e U
(\delta_{a[b}\Psi_{cd]e}-\delta_{e[b}\Psi_{cd]a})\nonumber \\
&& + \frac{a_L^*a_R^*e^{-3U}}{3a_-^2}\Big[\frac{-3}{8}F^{efg}_{\ \
\ a}
 (\Psi\wedge \bar\Psi)_{bcdefg}-12F^e_{\ a[bc}J_{d]e}+\frac{9i}{2}
F^{ef}_{\ \ a[b}(J\wedge
  J)_{cd]ef} \nonumber \\ && +
6iF_{abcd}\Big]
 \Big(\partial_a U +
\frac{2}{a_-^2}(-a_L^*\partial_a a_L+ a_R^* \partial_a a_R ) \Big)
\Psi_{bcd} \nonumber \\ &&
 + \frac{2}{a_-^2}(-a_L^*\partial_a
a_R^*+a_R^*\partial_a a_L^*)(v\wedge J)_{bcd} + \frac{2 }{a_-^2}(W
a_L^*a_R + \bar W a_L a_R^*)(v\wedge  \Psi)_{abcd} \nonumber
\\&& -\frac{1 }{a_-^2}( W a_L^{*2} + \bar W a_R^{*2} )[i(J\wedge
J)_{abcd}+6\delta_{a[b}J_{cd]}] \label{399}
\end{eqnarray}
and hence
\begin{eqnarray}
 (d\Psi)_{abcd} & = & -6e^{-3U} F^{fg}_{\ \ [ab}(v \wedge
\Psi)_{cd]fg} +\frac{ime^{-3U}}{24}(v\wedge  \Psi)_{abcd}\nonumber \\
&& + \frac{4a_L^*a_R^*e^{-3U}}{3a_-^2}\Big[\frac{-3}{8}F^{efg}_{\
\ \ [a}(\Psi\wedge \bar\Psi)_{bcd]efg}-12F^e_{\
[abc}J_{d]e}+\frac{9i}{2}F^{ef}_{\ \ [ab}(J\wedge
J)_{cd]ef} \nonumber \\ &&
+6iF_{abcd}\Big]
 + 4\Big(7\partial_{[a} U + \frac{2}{a_-^2}(-a_L^*\partial_{[a}
a_L+ a_R^* \partial_{[a} a_R ) \Big) \Psi_{bcd]} \nonumber \\ && +
\frac{8}{a_-^2}(-a_L^*\partial_{[a} a_R^*+a_R^*\partial_{[a}
a_L^*)(v\wedge J)_{bcd]} + \frac{8 }{a_-^2}(W a_L^*a_R + \bar W
a_L a_R^*)(v\wedge  \Psi)_{abcd} \nonumber \\ &&  -\frac{4i
}{a_-^2}( W a_L^{*2} + \bar W a_R^{*2} )(J\wedge J)_{abcd}
\label{500}
\end{eqnarray}
The superpotential terms can be written as
\begin{eqnarray}
(a_L^* a_R W+a_La_R^* \bar W)&=& \frac{i e^{-3U}}{36}[a_La_R(H
\haken \bar \Psi)-a_L^*a_R^*(H \haken \Psi)]\nonumber
\\&& -a_+^{2}\Big(\frac{1}{2}\partial U \haken v -
\frac{1}{288}e^{-3U} G\haken (J\wedge J)\Big) \label{507}\\
(a_L^{*2} W+a_R^{*2} \bar W)&=& \frac{i e^{-3U}a_+^2}{36}(H \haken
\bar \Psi) \nonumber
\\&& -a_L^*a_R^*\Big(\frac{1}{2}\partial U \haken v -
\frac{1}{288}e^{-3U} G\haken (J\wedge J)\Big) \label{508}
\end{eqnarray}
which are from eq.\ (\ref{383}-\ref{384}).



%

\providecommand{\href}[2]{#2}\begingroup\raggedright\endgroup

\end{document}